\documentclass[journal,draftcls,onecolumn,12pt,twoside]{IEEEtranTCOM}

\usepackage{exscale}
\usepackage{amsmath,esint}
\usepackage{amssymb}
\usepackage{graphicx,float}
\usepackage{fixltx2e}
\usepackage{subcaption}
\usepackage{epstopdf}
\usepackage{multirow}
\usepackage{array}
\usepackage{cite}
\usepackage{color}

\DeclareMathOperator*{\vc}{vec}
 
\newcommand{\bv}{\mathbf{v}}    % transmit signal vector
\newcommand{\bi}{\mathbf{i}}    % transmit signal current
\newcommand{\bx}{\mathbf{x}}    % transmit signal vector
\newcommand{\bs}{\mathbf{s}}    % transmit signal vector
\newcommand{\bw}{\mathbf{w}}    % weight vector
\newcommand{\bg}{\mathbf{g}}    % weight vector
\newcommand{\bH}{\mathbf{H}}    % channel matrix
\newcommand{\bG}{\mathbf{G}}    % fading path gain
\newcommand{\bn}{\mathbf{n}}    % noise vector
\newcommand{\bh}{\mathbf{h}}    % channel vector
\newcommand{\bN}{\mathbf{N}}    % noise covariance
\newcommand{\bC}{\mathbf{C}}    % coupling matrix
\newcommand{\bD}{\mathbf{D}}    % coupling matrix
\newcommand{\bZ}{\mathbf{Z}}    % impedance matrix
\newcommand{\bW}{\mathbf{W}}    % impedance matrix
\newcommand{\bI}{\mathbf{I}}    % Current vector or identity matrix
\newcommand{\bV}{\mathbf{V}}    % voltage vector
\newcommand{\bF}{\mathbf{F}}    % matched channel matrix
\newcommand{\bu}{\mathbf{u}}    % weight vector
\newcommand{\bS}{\mathbf{S}}    % covariance
\newcommand{\bT}{\mathbf{T}}    % normalized matched channel matrix
\newcommand{\bR}{\mathbf{R}}    % resistance matrix or spatial correlation
\newcommand{\bX}{\mathbf{X}}    % reactance matrix
\newcommand{\bY}{\mathbf{Y}}    % reactance matrix
\newcommand{\bU}{\mathbf{U}}    % auxiliary matrix
\newcommand{\bA}{\mathbf{A}}    % auxiliary matrix
\newcommand{\bB}{\mathbf{B}}    % auxiliary matrix
\newcommand{\bQ}{\mathbf{Q}}    % auxiliary matrix
\newcommand{\define}{\triangleq}    % Bold vector b

\newcommand{\mbbC}{\mathbb{C}}	% field of complex number
\newcommand{\mccn}{\mathcal{CN}}	% 	complex Gaussian

\newcommand{\bLam}{\mathbf{\Lambda}}

\newcommand{\bSig}{\mathbf{\Sigma}}

\newcommand{\btt}{\bsym{\theta}}

\newcommand{\bzro}{\mathbf{0}}

\newcommand{\ben}{\begin{enumerate}} 	  	% 	Begin Enumerate
    \newcommand{\een}{\end{enumerate}} 			% 	End Enumerate
\newcommand{\beq}{\begin{equation}} 	  	% 	Begin Equation
    \newcommand{\eeq}{\end{equation}} 			% 	End Equation
\newcommand{\bes}{\begin{equation*}}
    \newcommand{\ees}{\end{equation*}}

\newcommand{\bea}{\begin{eqnarray}}		% 	Begin Equation Array
    \newcommand{\eea}{\end{eqnarray}} 		% 	End Equation Array
\newcommand{\beas}{\begin{eqnarray*}}
    \newcommand{\eeas}{\end{eqnarray*}}
\newcommand{\ba}{\begin{array}}
    \newcommand{\ea}{\end{array}}
\newcommand{\sbea}{\nopagebreak[3]\samepage\begin{eqnarray}}
    \newcommand{\seea}{\end{eqnarray}\pagebreak[0]}
\newcommand{\sbeas}{\nopagebreak[3]\samepage\begin{eqnarray*}}
    \newcommand{\seeas}{\end{eqnarray*}\pagebreak[0]}

\newcommand{\er}[1]{{\rm(\ref{#1})}}

\newcommand{\bit}{\begin{itemize}}
    \newcommand{\eit}{\end{itemize}}
\newcommand{\bsym}{\boldsymbol}

\newcommand{\nn}{\nonumber}

\DeclareMathOperator{\Trace}{Tr}
\DeclareMathOperator{\diag}{diag}
\DeclareMathOperator{\Real}{Re}

\newtheorem{theorem}{Theorem}%[section]
\newtheorem{lemma}{Lemma}

\newtheorem{corollary}{Corollary}

\newenvironment{proof}[1][Proof]{\begin{trivlist}
        \item[\hskip \labelsep {\bfseries #1}]}{\end{trivlist}}

\newcommand{\qed}{\nobreak \ifvmode \relax \else
    \ifdim\lastskip<1.5em \hskip-\lastskip
    \hskip1.5em plus0em minus0.5em \fi \nobreak
    \vrule height0.75em width0.5em depth0.25em\fi}

\def\figurename{Fig.}
\def\iflatex{\iftrue}
\def\ifcomments{\iffalse}

\begin{document} 
  \title{Antenna Impedance Estimation at MIMO Receivers}
  \author{
      Shaohan~Wu~and~Brian~L.~Hughes\footnote{S. Wu is with MediaTek USA, Irvine CA, 92606 (e-mail: shaohan.wu@mediatek.com), and B. L. Hughes is with the Department of Electrical and Computer Engineering, North Carolina State University, Raleigh, NC 27695 (e-mail: blhughes@ncsu.edu).}
%    Department of Electrical and Computer Engineering \\
%    North Carolina State University \\
%    Raleigh, NC 27695-7911 \\
%    \{\em swu10,blhughes\}@ncsu.edu
%\IEEEauthorblockN{Shaohan~Wu}\\
%  \IEEEauthorblockA{
%    {MediaTek USA Inc.}, Irvine, CA \\
%    {shaohan.wu}@mediatek.com}\\
%  \and
%  \IEEEauthorblockN{Brian~L.~Hughes}\\
%  \IEEEauthorblockA{
%%      {Department of Electrical and Computer Engineering} \\
%    {North Carolina State University}, Raleigh, NC \\
%    {blhughes}@ncsu.edu}
}

  %\markboth{IEEE Transactions on Communications}{DRAFT}
  \markboth{}{DRAFT}
  \date{\today}
  \maketitle
  \begin{abstract}
%    \blfootnote{This material is based
%      upon work supported by the National Science Foundation under Grant
%      1343309. Any opinions, findings, and conclusions or
%      recommendations expressed in this material are those of the
%      author(s) and do not necessarily reflect the views of the National
%      Science Foundation.}
    
    This paper considers antenna impedance estimation based on training sequences at MIMO receivers. The goal is to firstly leverage extensive resources available in most wireless systems for channel estimation to estimate antenna impedance in real-time. We assume the receiver switches its impedance in a predetermined fashion during each training sequence. Based on voltage observation across the load,  a
    classical estimation framework is developed incorporating the Rayleigh fading assumption. We then derive in closed-form a maximum-likelihood (ML) estimator  under i.i.d. fading and show this same ML estimator is a method of moments (MM) estimator in correlated channels.  Numerical results suggest a fast algorithm, i.e., MLE in i.i.d. fading and the MM estimator in correlated fading, that estimates the unknown antenna impedance in real-time for all Rayleigh fading channels.
  \end{abstract}
  
  \begin{IEEEkeywords}
    Antenna Impedance Estimation, Maximum-Likelihood Estimator, MIMO, Training Sequences.
  \end{IEEEkeywords}

%%%%%%%%%%%%%%%%
%	transitional language
%%%%%%%%%%%%%%%% 

\section{Introduction}

 Over the past two decades, several works have demonstrated that impedance matching between the receive antenna and front-end significantly impacts channel capacity in wireless channels\cite{domi,domi2,lau,wall,gans,gans2}. In order to implement capacity-optimal matching, the receiver must know the antenna impedance. However, this impedance depends on time-varying near-field loading conditions and often changes in an unpredictable manner. To mitigate such variation, researchers have proposed antenna impedance estimation techniques\cite{wu,wu2,wu2021,wu_PCA,wu3,ali,hass,moha,vasi}. 

In previous works,  impedance estimation at  single-antenna receivers has been studied \cite{wu,wu2,wu_PCA, wu_cgr}. However, modern receivers are often equipped with multiple antennas for multiplexing and/or diversity benefits. Therefore, in this paper, we investigate the general and more important problem of estimating the antenna impedance matrix at MIMO receivers. Hassan and Wittneben investigated joint MIMO impedance and channel estimation using least squares\cite{hass}, and Wu solved a similar problem using hybrid estimation\cite{wu2021}. However, it remains unclear if either of these aforementioned approaches leads to the optimal MIMO impedance estimator. We fill in this gap in this paper. 

This paper considers antenna impedance estimation algorithms using training data for multiple-input, multiple-output (MIMO) communication systems. We assume the receiver switches its impedance in a predetermined fashion during each training sequence. In i.i.d. Rayleigh fading channels,  the maximum-likelihood (ML) estimator is derived for the impedance matrix as a function of the top block eigen-vector of the sample covariance matrix. This ML estimator is shown to be a  method of moments (MM) when the fading channel is temporally correlated. Fundamental lower bounds, e.g., Cram\'er-Rao bounds (CRB), on these estimators are derived and important properties of these estimators, e.g., bias and efficiency, are explored through numerical simulations.  The trade-off between channel and impedance estimation is  demonstrated empirically. 

The rest of the paper is organized as follows. We present our system model in Sec.~\ref{4secII}, derive a set of maximum-likelihood estimators for the MIMO antenna impedance and channel covariance matrix in Sec.~\ref{4secIII}, and derive  method of moments (MM) estimators of these matrices under multiple packets scenarios and discuss ML estimators in Sec.~\ref{4secIV}.  We then explore important properties of the estimators through numerical examples in Sec.~\ref{4secV}, and summarize our conclusions in Sec.~\ref{4secVI}.

%\newpage
\section{System Model}\label{4secII}
Consider a narrowband multiple-input multiple-output (MIMO) communications link with $M$ receive  antennas and $N$ transmit  antennas. The receiver  model is illustrated in Fig.~\ref{fig_SystemModel_MIMO}. This  circuit model is identical to the ones widely used  to model a scenario, where amplifier noise dominates \cite{gans2,lau,wall}.  This model is also a special case of the more general and complex models, which include additional noise sources, e.g., sky-noise and downstream noise \cite{domi,domi2,ivrl}.

\iflatex
\begin{figure*}[t!]
	\begin{center}
		\includegraphics[width=0.6\textwidth, keepaspectratio=true]{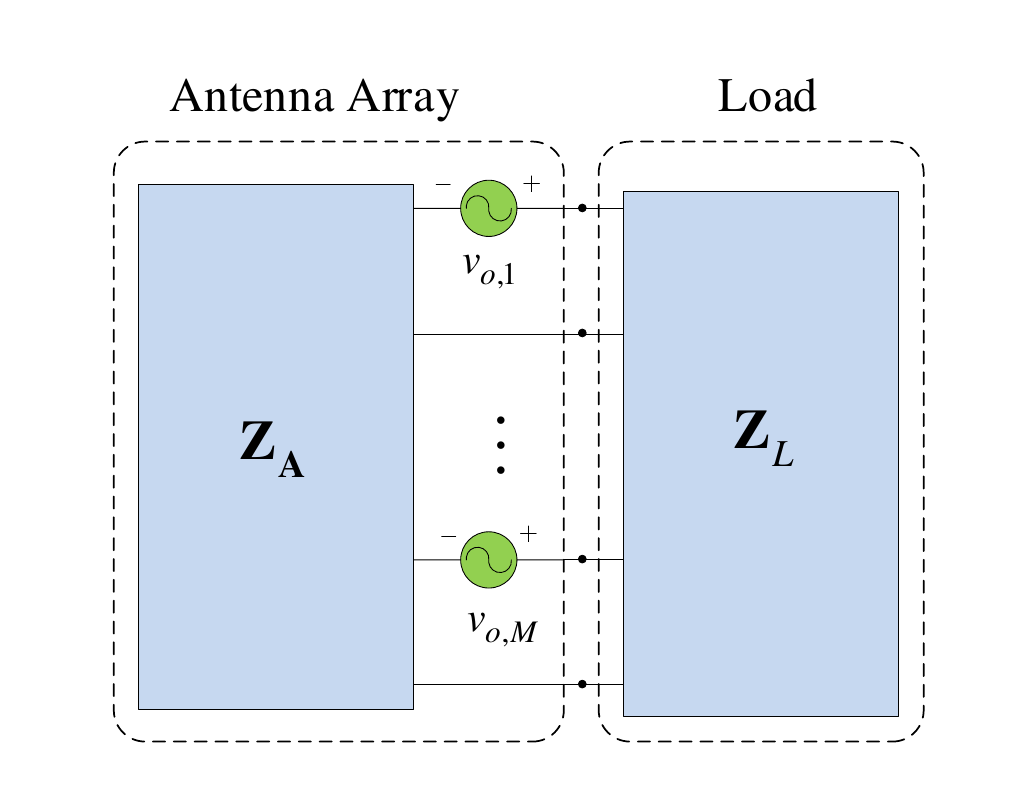}
	\end{center}
	\vspace{-12pt}
	\caption{Circuit model of a multiple-antenna receiver}
	\vspace{-18pt}
	\label{fig_SystemModel_MIMO}
\end{figure*}
\fi

In Fig.~\ref{fig_SystemModel_MIMO}, we model the antenna array by its Thevenin equivalent,
\beq\label{antenna_MIMO}
\bv \ = \ \bZ_\bA \bi + \bv_o \ , 
\eeq
where $\bv, \bi \in \mathbb{C}^M$ are the voltage across, and current into, the antenna array terminals. In particular, the antenna impedance is a symmetric matrix in $\mbbC^{M\times M}$ due to the reciprocity theorem of electromagnetics \cite{bala}, 
\beq\label{impd}
\bZ_\bA \ = \ \bR_\bA + j \bX_\bA \ ,
\eeq
where $\bR_\bA$ and $\bX_\bA$ are the resistance and reactance matrices, respectively. The incident electromagnetic field induces open-circuit voltage $\bv_o \in \mathbb{C}^M$ in \er{antenna_MIMO}.  Under flat-fading conditions, the open-circuit voltage $\bv_o$ is modeled as \cite{domi2}
\beq\label{oc_volt}
\bv_o \ = \ \bG\bx \ ,
\eeq
where $\bx\in\mbbC^N$ is the transmitted symbol and $\bG\in\mbbC^{M\times N}$ is the matrix of fading path gains. Similar to the previous two papers, we consider a Rayleigh fading environment, where transmit antennas are sufficiently separated.  Thus, columns of $\bG$ are modeled as i.i.d. zero-mean, complex Gaussian random vectors, $\bg_i\sim\mccn(\bzro, \bSig_\bg)$. 
As shown in Fig.~\ref{fig_SystemModel_MIMO}, noisy voltage signal  $\bv_L\in\mbbC^M$ across load impedance $\bZ_L\in\mbbC^{M\times M}$ is observed \cite{gans2,lau,wall}, 
\beq
\bv_L \ = \ \bZ_L\left({\bZ_\bA + \bZ_L}\right)^{-1}\bG\bx+\bn_L  \ , 
\eeq
where the noise $\bn_L\in\mbbC^M$ is a zero-mean, circularly-symmetric, complex Gaussian random vector with covariance $E[\bn_L\bn_L^H]=\bSig_L$, which is hereafter denoted by $\bn_L\sim \mccn(\bzro,\bSig_L)$. 

As mentioned in the previous paper,   performance of estimators typically depend on the signal-to-noise ratio (SNR) in estimation theory, which is conventionally defined, for example, as $\Trace[\bv_L\bv_L^H\bSig_L^{-1}]$ \cite[Sec.~II-A]{domi2}. 
In circuit theory, however, power depends on both voltage and current \cite[eq.~22]{ivrl}. 
This estimation-theory SNR formula does not correctly predict the ratio of the physical signal power and noise power in the receiver front-end. For a given $\bv_o$, this ratio of  physical signal power to noise power across the load is given by 
\beq
\rho \ \define \  \frac{1}{\sigma_{n}^2}\Trace\left[\bR_L^{1/2}\left(\bZ_\bA + \bZ_1\right)^{-1} \bv_o\bv_o^H\left(\bZ_\bA + \bZ_1\right)^{-H}\bR_L^{1/2}\right] \ ,
\eeq
where $\sigma_n^2$ represents the noise power at the output of the amplifier and $\bR_L=\Real\{\bZ_L\}$ is the load resistance. As in the previous paper, we correct this discrepancy by defining $\bSig_L$ in a way that ensures the SNR and physical power ratio $\rho$ coincide:
\[
\bSig_L\ \define \sigma_n^2 \bZ_L\bR_L^{-1} \bZ_L^H\ .
\]
With this definition, it is convenient to redefine the observed signal as
\beq\label{sgl_model_MIMO}
\bw \ \define \ \bR_L^{1/2}\bZ_L^{-1}\bv_L \ = \ {\bR_L^{1/2}}\left({\bZ_\bA + \bZ_L}\right)^{-1}\bG\bx+\bn  \ , 
\eeq
where $\bn\sim\mccn(\bzro,\sigma_n^2\bI_M)$  represents physical noise power referred to the amplifier output. 
This signal model \eqref{sgl_model_MIMO} correctly connects the estimator performance to physical signal-to-noise power ratio. As mentioned before in the previous paper, this connection is essential to accurately predict the impact of impedance mismatch at MIMO receivers on important system-level metrics, such as channel capacity. 

Suppose the channel gain matrix and antenna impedance matrix are unknown to the receiver. As in the previous two papers, our objective is to jointly estimate these two matrices using observations of known training sequences. 
%Suppose the transmitter is a base station, with $N$ uncoupled transmit antennas. 
Suppose the transmitter sends a known training sequence of length $T$, i.e., $\bx_1, \dots, \bx_T \in\mbbC^N$ to the receiver, during which the receiver synchronously shifts its impedance as $\bZ_{L,1}, \dots, \bZ_{L,T}$. Also assume both the fading path gain $\bG$ and impedance $\bZ_A$ remain fixed during each transmission. The received observations take the following form,
\beq
\bw_t \ = \ {\bR_{L,t}^{1/2}}\left({\bZ_\bA + \bZ_{L,t}}\right)^{-1}\bG\bx_t+\bn_t  \ , 
\eeq
where $t=1,2,\dots,T$ and the additive noises $\bn_t\sim\mccn(\bzro,\sigma_n^2\bI_M)$ are independent and identically distributed (i.i.d.). 

We again assume the load impedance takes on two possible matrices, 
\beq\label{imp_shift_MIMO}
\bZ_L \ = \ \begin{cases}
	\bZ_1 \ ,& 1\leq t\leq K  \ ,\\
	\bZ_2 \ , & K\leq t \leq T \ .
\end{cases}
\eeq
Similar to the previous two papers, we assume $\bZ_L=\bZ_1$ is the load impedance used to receive the transmitted data, and is matched to our best estimate of $\bZ_A$; additionally $\bZ_L=\bZ_2$ is an impedance variation introduced in order to make $\bZ_A$ observable. To estimate $\bZ_A$,  $\bZ_1\neq \bZ_2$ is required. 

Note that in order the perform optimal detection of the transmitted symbols in \er{sgl_model_MIMO}, an accurate estimate of the entire matrix coefficient of $\bx$ is ideal, not simply the fading path gain matrix $\bG$. This motivates the definition of 
an effective channel matrix that communication algorithms need,
\beq\label{H_MIMO}
\bH \ \define \ \bR_1^{1/2}\left(\bZ_\bA + \bZ_1\right)^{-1}\bG\in\mbbC^{M\times N}\ ,
\eeq
whose columns  are also i.i.d. zero-mean, complex Gaussian, $\bh_i\sim\mccn(\bzro,\bSig_\bh)$, and
\beq\label{Sig_h_MIMO}
\bSig_\bh \ = \  \bR_1^{1/2}\left(\bZ_\bA + \bZ_1\right)^{-1} \bSig_\bg  \left(\bZ_\bA + \bZ_1\right)^{-H}\bR_1^{1/2} \ .
\eeq
With this choice of load impedance in \er{imp_shift_MIMO} and definition of $\bH$ \er{H_MIMO}, we can express the observations in a simpler, bilinear form.
The voltage observation \er{sgl_model_MIMO} at the load is then, 
\beq\label{4observations}
\bw_t \ = \ \begin{cases}
	\bH \bx_t + \bn_{t} \ , & 1\leq t \leq K \\
	\bF \bH \bx_t + \bn_{t} \ , & K+1\leq t \leq T 
\end{cases}
\eeq
where $\bn_t\sim\mccn(\bzro,\sigma_n^2\bI_M)$ are independent over time $t$ and we define $\bF\in\mbbC^{M\times M}$ as a one-to-one mapping of $\bZ_\bA$ for mathematical convenience, conditioned on $\bZ_1\neq\bZ_2$, 
\bea\label{F_MIMO}
\bF &=& \bR_2^{1/2}\left(\bZ_2 + \bZ_\bA\right)^{-1}\left(\bZ_1 + \bZ_\bA\right)\bR_1^{-1/2}\ .
%&=& \left(\bI_M+\bZ_\bA\bZ_2^{-1}\right)^{-1}\left(\bI_M+\bZ_\bA\bZ_1^{-1}\right) \ .
\eea

We present this paper as a generalization of the previous paper to MIMO receivers. Here the goal  is to derive maximum-likelihood (ML) estimators for $\bH, \bZ_A$ and $\bSig_\bh$ based on the observations \er{4observations}. 
From the invariance principle of maximum-likelihood  estimation (MLE) \cite[pg. 185]{kay}, knowing the MLE of $\bF$ is equivalent to knowing that of $\bZ_A$ and vice versa. Theoretically it suffices to derive estimators for $\bH, \bF$ and $\bSig_\bh$. Specifically, we follow the two-step procedure as described in the last paper: First, we consider joint maximum-likelihood estimation of $\bF$ and $\bSig_\bh$, treating $\bH$ as a nuisance parameter. Second, given estimates of $\bF$ and $\bSig_\bh$, we then estimate $\bH$ using minimum mean-squared error estimation. Again we focus exclusively on estimators for $\bF$ and $\bSig_\bh$ in the next two sections; estimators for $\bH$ will be explored through numerical examples in Sec.~\ref{4secV}.

\section{Maximum-Likelihood Estimators}\label{4secIII}

In this section, we derive maximum-likelihood (ML) estimators for $\bF$ and $\bSig_\bh$ based on observations in \er{4observations}. It is often convenient to find sufficient statistics before deriving the ML estimators.

We write \eqref{4observations} in matrix form, after defining $\bX_1\define[\bx_1,\bx_2,\dots,\bx_K]\in\mbbC^{N\times K}$ and $\bX_2\define[\bx_{K+1},\bx_{K+2},\dots,\bx_T]\in\mbbC^{N\times (T-K)}$,
\bea\label{Ws}
\bW_1&=& \bH \bX_1 + \bN_1 \ , ~~~~
\bW_2 \ = \ 	\bF \bH \bX_2 + \bN_2 \ ,
\eea
where $\bN_1$ and $\bN_2$ are analogously defined, independent and have i.i.d. entries $\mccn(0,\sigma_n^2)$. 

The known training sequences for MIMO channel estimation are often equal-energy and orthogonal. We further assume $K=T/2$ and equal-energy and orthogonal training over the first  and last $K$ symbols i.e.,
\beq\label{Xs_def}
\bX_1\bX_1^H \ = \   \bX_2\bX_2^H \ = \ \frac{PK}{N}\bI_N  \ ,
\eeq
This can be achieved by using a normalized discrete Fourier transform (DFT) matrix, e.g., \cite[eq. 10]{bigu}. 
We present a sufficient statistic in the next lemma.

\begin{lemma}[Sufficient Statistic]\label{4lem_ss}	
	Consider the observations $\bW_1$ and $\bW_2$ defined in \er{Ws} and known training sequences in \er{Xs_def}. Then 
	\beq\label{Ys_MIMO}
	\bY_1 \ =\ \left(\frac{2N}{PT}\right) \bW_1\bX_1^H\ , ~~~~ \bY_2 \ = \ \left(\frac{2N}{PT}\right)\bW_2\bX_2^H\ , 
	\eeq
	are sufficient for estimating unknown matrices $\bF$ and $\bSig_\bh$.  Moreover, $\bY_1 - \bH$ and $\bY_2 -\bF\bH$ are independent random matrices with i.i.d. $\mccn(0,\sigma^2)$ entries, where $\sigma^2\define 2N\sigma_n^2/PT$. 
	$\hfill\diamond$
\end{lemma}
\begin{proof}
	From \er{Ws} and \er{Xs_def}, we have $\bY_1 = \bH + \left(\frac{2N}{PT}\right)\bN_1\bX_1^H$. To show the entries of the last matrix are i.i.d., we vectorize it,
	\beq
	\left(\frac{2N}{PT}\right) \vc \left(\bN_1\bX_1^H\right) \ = \  \left(\frac{2N}{PT}\right) \left(\bX_1^*\otimes \bI_M\right) \vc \bN_1 \in\mbbC^{MN} \ ,
	\eeq
	which is zero-mean and has covariance matrix  $\left(\frac{2N}{PT}\right)^2 \left(\bX_1^*\bX_1^T\otimes \bI_M\right) \sigma_n^2\bI_{MN} =\sigma^2\bI_{MN}$, where by definition $\sigma^2= 2N\sigma_n^2/PT$.  Note an identity of Kronecker product 
	$\vc \left(\bA\bB\bC\right) = \left(\bC^T\otimes\bA\right)\vc \bB$ is used \cite{brew}. This shows that $\bY_1 - \bH$ is a random matrices with i.i.d. $\mccn(0,\sigma^2)$ entries. Similarly, $\bY_2 -\bF\bH = \left(\frac{2N}{PT}\right)\bN_2\bX_2^H$ is also a random matrices with i.i.d. $\mccn(0,\sigma^2)$ entries. The independence between these two matrices follow from that noises are independent over time \er{Ws}. 
	
	From the Neyman-Fisher theorem \cite[pg. 117]{kay}, to prove sufficiency of \er{Ys_MIMO} it suffices to show that $p\left(\bW_1,\bW_2;\bF,\bSig_\bh\right)$ factors into a product $g\left(\bY_1,\bY_2, \bF,\bSig_\bh\right) f\left(\bW_1,\bW_2\right)$, where $f$ does not depend on $\bY_1,\bY_2, \bF,\bSig_\bh$ and $g$ does not depend on $\bW_1,\bW_2$. 
	We prove this using the conditional pdf
	\beq
	p\left(\bW_1,\bW_2;\bF,\bSig_\bh\right) \ = \ E_\bH\left[p\left(\bW_1,\bW_2|\bH;\bF,\bSig_\bh\right)\right]\ ,
	\eeq
	where the expectation $E_\bH[\cdot]$ is with respect to $\bH$ \er{H_MIMO}. Since $\bW_1$ and $\bW_2$ are conditionally independent given $\bH$, we have
	\bea\label{4lem_ss_pdf}
	&&(\pi\sigma_n^2)^{NT} p\left(\bW_1,\bW_2;\bF,\bSig_\bh\right) \nn\\
	 &=&  E_\bH\left[\exp\left(-\frac{1}{\sigma_n^2}\left\lVert\bW_1-\bH\bX_1\right\rVert^2-\frac{1}{\sigma_n^2}\left\lVert\bW_2-\bF\bH\bX_2\right\rVert^2\right)\right]\nn\\
%	&=&E_\bH\left[\exp\left(\frac{\Trace\left[\bW_1^H\bH\bX_1\right]}{\sigma_n^2}+\frac{\Trace\left[\bX_1^H\bH^H\bW_1\right]}{\sigma_n^2}-\frac{\Trace\left[\bX_1^H\bH^H\bH\bX_1\right]}{\sigma_n^2}\right. \right.\nn\\
%	&&\left. \left. + \frac{\Trace\left[\bW_2^H\bF\bH\bX_2\right]}{\sigma_n^2}+ \frac{\Trace\left[\bX_2^H\bH^H\bF^H\bW_2\right]}{\sigma_n^2}-\frac{\Trace\left[\bX_2^H\bH^H\bF^H\bF\bH\bX_2\right]}{\sigma_n^2}\right)\right]\nn\\
%	&&\exp\left(-\frac{1}{\sigma_n^2}\left\lVert\bW_1\right\rVert^2-\frac{1}{\sigma_n^2}\left\lVert\bW_2\right\rVert^2\right) \nn\\
	&=& E_\bH\left[\exp\left(\frac{2\Real\Trace\left[\bH^H\bW_1\bX_1^H\right]}{\sigma_n^2}-\frac{\Trace\left[\bH^H\bH\bX_1\bX_1^H\right]}{\sigma_n^2}+ \frac{2\Real\Trace\left[\bH^H\bF^H\bW_2\bX_2^H\right]}{\sigma_n^2}\right. \right.\nn\\
	&& \left. \left. -\frac{\Trace\left[\bH^H\bF^H\bF\bH\bX_2\bX_2^H\right]}{\sigma_n^2}\right)\right] \exp\left(-\frac{1}{\sigma_n^2}\left\lVert\bW_1\right\rVert^2-\frac{1}{\sigma_n^2}\left\lVert\bW_2\right\rVert^2\right) \nn\\
	&=&E_\bH\left[\exp\left(\frac{2\Real\Trace\left[\bH^H\bY_1+\bH^H\bF^H\bY_2\right]}{\sigma^2}-\frac{\Trace\left[\bH^H\bH+\bH^H\bF^H\bF\bH\right]}{\sigma^2}\right) \right]\nn\\
	&&\exp\left(-\frac{1}{\sigma_n^2}\left\lVert\bW_1\right\rVert^2-\frac{1}{\sigma_n^2}\left\lVert\bW_2\right\rVert^2\right)\ ,
	\eea
	where $\lVert\bA \rVert^2=\Trace[\bA^H\bA]$ denotes the Frobenius norm. Also, the third equality follows from the identities $2\Real\Trace[\bA] = \Trace[\bA] + \Trace[\bA^H]$ and $\Trace[\bA\bB] = \Trace[\bB\bA]$, and the fourth equality follows from \er{Xs_def} and the definition of $\sigma^2$. 
	In \er{4lem_ss_pdf}, denote the first factor by $(\pi\sigma_n^2)^{NT}g\left(\bY_1,\bY_2, \bF,\bSig_\bh\right)$ and the second by $ f\left(\bW_1,\bW_2\right)$. Note $g$ depends on 
	$\bY_1,\bY_2, \bF$ and $\bSig_\bh$ (through the expectation) but not on $\bW_1,\bW_2$. And $f$ depends on $\bW_1,\bW_2$ only, not $\bY_1,\bY_2, \bF,\bSig_\bh$. This completes the proof. 
\end{proof}

Based on the sufficient statistics in \er{Ys_MIMO}, we want to estimate the following complex parameters
\beq\label{theta_MIMO}
\btt \ \define \vc \begin{bmatrix}
	\bF & \bSig_\bh
\end{bmatrix} \ ,
\eeq
where $\bF$ is defined in \er{F_MIMO} and $\bSig_\bh$ in \er{Sig_h_MIMO}. Here we present the maximum-likelihood (ML) estimator, such that
\beq
\hat{\btt}_{ML} \ \define \ \arg\max_{\btt} p\left(\bY_1,\bY_2;\btt\right) \ .
\eeq
The next theorem shows the ML estimator can be calculated via block eigen-decomposition using the sufficient statistic given in \er{Ys_MIMO}. 

\begin{theorem}[Single-Packet ML Estimators]\label{MLE_ch4}
	Let $\bY_1$ and $\bY_2$ be the sufficient statistics in \er{Ys_MIMO}. Suppose $\bF$ and $\bSig_\bh$ are unknown. Consider the sample covariance matrix,
	\beq\label{S_MIMO}
	\bS \ \define \ \frac{1}{N} \begin{bmatrix}
		\bY_1\bY_1^H &  \bY_1\bY_2^H\\
		\bY_2\bY_1^H & \bY_2\bY_2^H
	\end{bmatrix} \in\mbbC^{2M\times 2M}\ . 
	\eeq
	The eigen-decomposition of $\bS$ can be written as
	\beq
	\bS \bU_\bs \ = \ \bU_\bs \diag(\mu_1,\dots,\mu_{2M}) \ ,
	\eeq
	where $\diag(\cdot)$ denotes a square diagonal matrix with its input as diagonal entries, and the eigen-values $\mu_k\geq 0$ are in descending order. Define the unitary eigen-vector matrix $\bU_\bs$ as a 2 by 2 block matrix, i.e., 
	\beq
	\bU_\bs \ \define \ \begin{bmatrix}
		\bU_{\bs11} & \bU_{\bs12} \\
		\bU_{\bs21} & \bU_{\bs22}
	\end{bmatrix} \ , 
	\eeq
	where $\bU_{\bs ij}\in\mbbC^{M\times M}$ and $i,j=1,2$. 
	Then, the maximum-likelihood estimate of $\btt$ is,
	\beq
	\hat{\btt}_{ML} \define \vc \left[\hat{\bF}_{ML} ~~ \hat{\bSig_\bh}\right] \ ,
	\eeq
	where $\sigma^2\define 2N\sigma_n^2/PT$ and, conditioned on $\bU_{\bs11}$ is non-singular, 
	\beq
	\hat{\bF}_{ML}  \ = \  \bU_{\bs21} \bU_{\bs11}^{-1} \ , ~~~~ \hat{\bSig_\bh} \ = \  \bU_{\bs11}\left(\diag(\mu_1,\dots,\mu_M) - \sigma^2\bI_M\right)^+ \bU_{\bs11}^H \ .
	\eeq
	Here $(\cdot)^+$ is an element-wise operator on real matrices, such that $[(\bA)^+]_{ij}\define \max\{[\bA]_{ij},0\}$.  
\end{theorem}

\begin{proof}
Consider the sufficient statistic in \eqref{Ys_MIMO}, and define
\beq\label{V_MIMO}
\bV \ \define \ \begin{bmatrix}
	\bY_1\\
	\bY_2
\end{bmatrix} \ = \ \begin{bmatrix}
	\bH + \bN_1\\
	\bF\bH + \bN_2
\end{bmatrix}\ = \ \begin{bmatrix}
	\bh_1 + \bn_{1,1}& \cdots &\bh_N + \bn_{1,N}\\
	\bF\bh_1 + \bn_{2,1}& \cdots & \bF\bh_N + \bn_{2,N}
\end{bmatrix} \ .
\eeq
Due to uncoupled transmit antennas, the $N$ columns of $\bH$ are also i.i.d. zero-mean, circularly-symmetric, complex Gaussian random vectors, i.e., $\bh_k \sim\mccn\left(\bzro_M,\bSig_\bh\right)$ for all $1\leq k\leq N$, where $\bSig_\bh\define E\left[\bh_k\bh_k^H\right]$. Thus, the prior information for channel $\bH$ can be written as,
\beq\label{pdf_H_MIMO}
p(\bH) = \frac{1}{\det\left(\pi \bI_N\otimes\bSig_\bh\right)} \exp\left[-\left(\vc \bH\right)^H\left( \bI_N\otimes\bSig_\bh^{-1}\right)\vc \bH\right] \ .
\eeq

The covariance matrix of $\bV$ is defined as,
\bea\label{CovV}
\bSig & \define & \frac{1}{N} E[\bV\bV^H] \ = \ \begin{bmatrix}
	\bSig_\bh + \sigma^2\bI_M& \bSig_\bh \bF^H\\
	\bF\bSig_\bh & \bF\bSig_\bh\bF^H + \sigma^2\bI_M
\end{bmatrix} \ . 
\eea

Since $\bSig$ is Hermitian, it decomposes into the following block eigen-system \cite{pere},
\bea\label{block_eigen}
\bSig &=& \begin{bmatrix}
	\bI_M\\
	\bF
\end{bmatrix} \bSig_\bh \begin{bmatrix}
	\bI_M&\bF^H
\end{bmatrix} + \sigma^2\bI_{2M} \ = \ \bB_1\bD_1\bB_1^H  + \bB_2\bD_2\bB_2^H \ ,
\eea
where $\bD_i\in\mbbC^{M\times M}$ and $\bB_i\in\mbbC^{2M\times M}$ (for $i=1,2$) are the block eigenvalues and orthonormal block eigenvectors of $\bSig$, 
\bea
\bB_1 &=&
\begin{bmatrix}
	\bI_M\\
	\bF
\end{bmatrix} \bA_1^{-\frac{1}{2}} \ , ~~~ \bB_2 \ = \ \begin{bmatrix}
	-\bF^H\\
	\bI_M
\end{bmatrix} \bA_2^{-\frac{1}{2}} \ ,  \nn\\
\bD_1 &=& \bA_1^{\frac{1}{2}}\bSig_\bh\bA_1^{\frac{1}{2}} + \sigma^2\bI_M \ , ~~~ \bD_2 \ = \ \sigma^2\bI_M \ , 
\eea
and $\bA_1$ and $\bA_2$ are Hermitian matrices defined as,
\beq\label{As}
\bA_1 \ \define \ \bF^H\bF + \bI_M \ , ~~~ \bA_2  \ \define \ \bF\bF^H + \bI_M \ . 
\eeq
Note both $\bA_1$ and $\bA_2$ are positive definite; also $\bF^H\bA_2=\bA_1\bF^H$ and $\bF\bA_1=\bA_2\bF$. 

Since eigenvalues of $\bD_i$ are also eigenvalues of $\bSig$ \cite[Th. 1.1]{pere}, and observe from \eqref{block_eigen} that $\bD_2$ is already diagonal, we write down the (scalar) eigenvalue decomposition of $\bSig$,
\beq\label{Sig_eig}
\bSig \ = \ \bU \bLam\bU^H \ \define \ \begin{bmatrix}
	\bB_1 \bQ & \bB_2
\end{bmatrix} \begin{bmatrix}
	\bLam_1 & \bzro_{M\times M}  \\
	\bzro_{M\times M} & \bLam_2 
\end{bmatrix} \begin{bmatrix}
	\bQ^H\bB_1^H\\
	\bB_2^H
\end{bmatrix}\ ,
\eeq
where $\bD_1\bQ = \bQ\bLam_1$,  $\bLam_1=\diag(\lambda_1,\dots,\lambda_M)$ is diagonal, $\bQ\in\mbbC^{M\times M}$ is unitary, and  $\bLam_2 = \sigma^2\bI_M$. Note $\lambda_j \geq \sigma^2$ for all $1\leq j \leq M$. In other words, the covariance matrix $\bSig$ has $2M$ real eigenvalues, where the largest $M$ eigenvalues are greater than or equal to $\sigma^2$, while the smallest $M$ eigenvalues are exactly  $\sigma^2$.

Under the assumptions above, we assume $\sigma^2$ is known but $\bLam_1$ and $\bU$ are unknown. If we find the joint ML estimator for $\bLam_1$ and $\bU$,  the ML estimators for $\btt$ \eqref{theta_MIMO} follows by the invariance principle of MLE \cite[Th. 7.4, pg. 185]{kay}. From \eqref{Sig_eig}, we firstly find the inverse of covariance  $\bSig$ using it scalar eigen system, 
\beq
\bSig^{-1} \ = \ \bU\bLam^{-1}\bU^H \ = \ \begin{bmatrix}
  \bB_1 \bQ & \bB_2
\end{bmatrix} \begin{bmatrix}
  \bLam_1^{-1} & \bzro_{M\times M}  \\
  \bzro_{M\times M} & \bLam_2^{-1} 
\end{bmatrix} \begin{bmatrix}
  \bQ^H\bB_1^H\\
  \bB_2^H
\end{bmatrix}  \ .
\eeq
We then write the density function of $\bV\define[\bv_1,\dots,\bv_N]$ in \eqref{V_MIMO} as,
\bea\label{likelihood}
&&p(\bV; \btt) \ = \  \det(\pi \bSig)^{-N} \exp \left(-\sum_{i=1}^{N}\bv_i^H\bSig^{-1}\bv_i\right) \nn\\
& = & \det(\pi \bSig)^{-N} \exp \left(-N\Trace[\bS\bSig^{-1}]\right) \nn\\
&= & \left(\pi \prod_{j=1}^{2M} \lambda_j\right)^{-N} \exp\left(-N \sum_{j=1}^{2M} \frac{\bu_j^H\bS\bu_j}{\lambda_j}\right) \nn\\
&=& \left(\pi \prod_{j=1}^{2M} \lambda_j\right)^{-N} \exp \left[N \sum_{j=1}^{M}\left(\frac{1}{\sigma^2}-\frac{1}{\lambda_j}\right)\bu_j^H\bS\bu_j  - \frac{N\Trace(\bS)}{\sigma^2}\right]  \ ,
\eea
where, in the second equality, sample covariance matrix defined in \er{S_MIMO} can be equivalently written in terms of $\bV=[\bv_1,\dots,\bv_N]$, 
\beq
\bS \ = \ \frac{1}{N} \bV\bV^H \ = \ \frac{1}{N} \sum_{i=1}^{N} \bv_i\bv_i^H\ ,
\eeq
and in the third equality we define $\bU\define[\bu_1,\dots,\bu_N]$. 
Since we have $\lambda_j\geq \sigma^2$ for $1\leq j \leq M$, the coefficients $\frac{1}{\sigma^2}-\frac{1}{\lambda_j}$ are non-negative and in decreasing order. The likelihood function is maximized, when 
$\bu_j$ is chosen to be the eigenvector corresponding to the $j$th largest eigenvalue of the sample covariance $\bS$. Assume the eigenvalues of $\bS$ are ordered in descending order, $[\mu_1,\dots,\mu_{2M}]$, the maximum of \eqref{likelihood} is,
\bea\label{likelihood_max}
\max_{\lambda_j, 1\leq j\leq M} p(\bV) & = & \left(\pi \prod_{j=1}^{2M} \lambda_j\right)^{-N} \exp \left(-N\sum_{j=1}^{M}\frac{\mu_j}{\lambda_j} - \frac{N}{\sigma^2}\sum_{k=M+1}^{2M}\mu_k\right) \nn\\
&\leq & \left(\pi \sigma^{2M}\prod_{j=1}^{M} \mu_j\right)^{-N} \exp\left(-NM- \frac{N}{\sigma^2}\sum_{k=M+1}^{2M}\mu_k\right)\ ,
\eea
where the last step is because $xe^{-x}$ is uniquely maximized at $x=1$, and equality is achieved by letting 
\beq
\hat{\lambda}_j = \max \{\mu_j, \sigma^2 \}\ , ~~~1\leq j \leq M \ .
\eeq 

Now it remains to show that \eqref{likelihood} can be achieved by actual estimates of $\bF$ and $\bSig_\bh$.  Assume that the $\bS\bU_\bs = \bU_\bs\diag(\mu_1,\dots,\mu_{2M})$, where 
\beq
\bU_\bs \ \define \ \begin{bmatrix}
	\bU_{\bs11}& \bU_{\bs12}\\
	\bU_{\bs21}&\bU_{\bs22}
\end{bmatrix} \ .
\eeq
Then we verify that the following estimates achieve the maximum in \eqref{likelihood},
\beq
\hat{\bSig} \ = \ \bU_\bs \diag(\hat{\lambda}_1,\dots,\hat{\lambda}_M,\sigma^2,\dots,\sigma^2)\bU_\bs^H \ . 
\eeq
Thus, the joint ML estimators for $\bF$ and $\bSig_\bh$ are, conditioned on  $\bU_{\bs11}$ is non-singular, 
\bea\label{mle_iid}
\hat{\bF}_{ML} &=& \bU_{\bs21} \bU_{\bs11}^{-1} \ , ~~~~ \hat{\bSig_\bh} \ = \  \bU_{\bs11}\left(\diag(\mu_1,\dots,\mu_M) - \sigma^2\bI_M\right)^+ \bU_{\bs11}^H \ .
\eea

	The derivation for $\hat{\bF}_{ML}=\bU_{\bs21} \bU_{\bs11}^{-1}$ is straightforward from \er{block_eigen}, as $\bF$ only depends on the eigen-vector matrix and not the eigen-values. 
	The derivation of $\hat{\bSig_\bh}$ follows also \er{block_eigen}, 
	\bea
	&&\hat{\bSig} -\sigma^2\bI_{2M}\ = \ \begin{bmatrix}
		\bI_M\\
		\hat{\bF}
	\end{bmatrix} \hat{\bSig_\bh} \begin{bmatrix}
		\bI_M&\hat{\bF}^H
	\end{bmatrix} \nn\\
	&&\begin{bmatrix}
		\bI_M&\hat{\bF}^H
	\end{bmatrix}\left(\hat{\bSig} -\sigma^2\bI_{2M}\right)\begin{bmatrix}
		\bI_M\\
		\hat{\bF}
	\end{bmatrix} \ = \ \left(\bI_M + \hat{\bF}^H\hat{\bF}\right)\hat{\bSig_\bh} \left(\bI_M + \hat{\bF}^H\hat{\bF}\right) \nn\\
	\hat{\bSig_\bh} & = & \left(\bI_M + \hat{\bF}^H\hat{\bF}\right)^{-1}\begin{bmatrix}
		\bI_M&\hat{\bF}^H
	\end{bmatrix}\left(\hat{\bSig} -\sigma^2\bI_{2M}\right)\begin{bmatrix}
		\bI_M\\
		\hat{\bF}
	\end{bmatrix}\left(\bI_M + \hat{\bF}^H\hat{\bF}\right)^{-1} \ ,
	\eea
	which reduces to \eqref{mle_iid} after plugging in $\hat{\bF}_{ML}$'s formula and some simplification. This completes the proof. 
\end{proof}

In order to study the efficiency of the joint ML estimators for $\bF$ and $\bSig_\bh$,   a natural next step is to find tight fundamental lower bounds on these estimators. Two fundamental lower bounds  are investigated next.

It has been shown that the (complex) Fisher information matrix (FIM) on $\btt$ is given by \cite[eq.~37]{wu3}, which extended results on real parameters to complex ones \cite[eq. 15.52]{kay}, 
\bea\label{FIM_MIMO}
&&[\boldsymbol{\mathcal{I}}(\btt)]_{ij} \ = \ N\Trace\left[\bSig^{-1}\frac{\partial \bSig}{\partial \theta_i^*}\bSig^{-1}\frac{\partial \bSig}{\partial \theta_j}\right] \ , 
\eea
where $\bSig$ is given in \er{block_eigen}. The error covariance matrix $\bC_{\hat{\btt}}$ of any unbiased estimator $\hat{\btt}$ is lower bounded by the Cram\'er-Rao bound (CRB), i.e.,  the inverse of $\boldsymbol{\mathcal{I}}(\btt)$,
\beq\label{crb_ch4}
\bC_{\hat{\btt}} \ \define \ E_{\bY_1,\bY_2;\btt}\left[\left(\hat{\btt}-\btt\right)\left(\hat{\btt}-\btt\right)^H\right] \ \geq \ \boldsymbol{\mathcal{I}}(\btt)^{-1}\ , 
\eeq
where the expectation is with respect to pdf in \er{likelihood} and $\bA\geq \bB$ means $\bA-\bB$ is positive semi-definite. 
The  formula in \er{crb_ch4} is unlikely to  simplify further without additional assumptions. Hence we evaluate the CRB numerically in  simulations. 

Another useful lower bound is the Miller-Chang bound (MCB) \cite{mill}. The formula for this bound is given below, but  details of its derivation are given in the Appendix of \cite{wu3}. It can be shown that square of Frobenius norm of  any unbiased estimator for $\bF$,  for all $\bH\in\mbbC^{M\times N}$ \eqref{H_MIMO}, is lower bounded by the MCB,
\bea\label{MCB}
\mathcal{M}(\bF) & = & \sigma^2\Trace \left(E\left[\left(\bH^*\bH^T\right)^{-1}\right]\right) \Trace \left(\bF\bF^H + \bI_M\right) \nn\\
&=&\begin{cases}
	\infty \ , & N \leq M \ ,\\
	\frac{\sigma^2}{N-M}\Trace\left(\bSig_\bh^{-1}\right) \Trace \left(\bF\bF^H + \bI_M\right) \ , & N > M \ ,
\end{cases} 
\eea
where  $N>M$ means  the number of antennas at the transmitter is greater than that at the receiver, and the final expression  follows from that $\bH^*\bH^T$ is a complex $M\times M$ Wishart matrix of degree $N$ \eqref{pdf_H_MIMO} and  the mean of its inverse is derived by Maiwald and Kraus \cite[eq. 39]{maiw}. Note the independence between columns of $\bH$ is essential. If $N\leq M$, the inverse mean of $\bH^*\bH^T$ is unbounded, then one packet is likely insufficient for any unbiased estimator of $\bF$ to have finite error in Frobenius norm. This motivates  finding estimators when observations from multiple packets are available.

\section{Estimators for Multiple Packets}\label{4secIV}
In the last section, we derived the maximum-likelihood estimators for $\bF$ and $\bSig_\bh$ using observations of training sequences from one packet. 
In this section, we consider estimators based on multiple packets, where channel varies from packet to packet.

As in the previous paper, suppose the transmitter sends $L$ identical training packets to the receiver. During transmission of each training packet, the receiver shifts it load impedance as described in \er{imp_shift_MIMO}. Similar to previous papers and last section, block fading is assumed, i.e., the channel remains constant within a packet but randomly varies from packet to packet. Similarly to \er{4observations} the signal observations at the $l$-th packet can be described as
\beq\label{4observations_mp}
\bw_{l, t} \ = \ \begin{cases}
	\bH_l \bx_t + \bn_{l,t} \ , & 1\leq t \leq K \\
	\bF \bH_l \bx_t + \bn_{l,t} \ , & K+1\leq t \leq T 
\end{cases}
\eeq
where the random noise vectors $\bn_{l,t}\sim\mccn(\bzro,\sigma_n^2\bI_M)$ are independent over packets $1\leq l\leq L$ and time $1\leq t\leq T$. We can express above observations in a compact matrix form, with a slight abuse of notation\footnote{Ideally, we would use notations like $\bW_{mp,1}$, $\bH_{mp}$, $\bN_{mp,1}$ and etc to distinguish them from their single-packet counterparts. We hereafter drop the subscript $mp$ for simplicity when confusion is unlikely to occur.}, 
\beq\label{Ws_mp}
\bW_{1} \ \define \ \bH(\bI_L\otimes \bX_1) + \bN_{1} \ ,  ~~\bW_{2} \ \define \ \bF\bH(\bI_L\otimes \bX_2) + \bN_{2} \ , 
\eeq
where $\bW_{1}\in\mbbC^{M\times LK}$, $\bW_{2}\in\mbbC^{M\times L(T-K)}$,   $\bX_1$ and $\bX_2$ are defined above \er{Ws}, $\bN_{1}$ and $\bN_{2}$ are independent random matrices each with i.i.d. entries $\mccn(0,\sigma_n^2)$,  and we define  the multi-packet channel as, again slightly abusing notation, 
\beq\label{eq_Hmp_def}
\bH \ \define  \ [\bH_1~ \cdots~\bH_L] \in\mbbC^{M\times NL} \ . 
\eeq
Here $\bH_l\in\mbbC^{M\times N}$ is the channel matrix for the $l$-th packet, whose columns are spatially i.i.d. complex Gaussian $\mccn(\bzro, \bSig_\bh)$  across transmit antennas but  temporally correlated across packets. If the normalized channel correlation is $\bC_\bh$, then the space-time correlation of $\bH$ is can be shown as 
\beq\label{vec_Hmp}
\vc \bH \sim\mccn\left(\bzro_{MNL}, \bC_\bh\otimes\bI_N\otimes\bSig_\bh\right) \ .
\eeq 
In this section, we assume $\bC_\bh$ is known. 

Similar to the previous section, the goal of this section is to derive estimators for both $\bF$ and $\bSig_\bh$, or $\btt$ as defined in \er{theta_MIMO}, treating $\bH$ as a nuisance parameter. Then, we explore estimators for $\bH$ given $\bF$ and $\bSig_\bh$ through numerical examples. The following lemma generalizes Lemma \ref{4lem_ss} to multiple packets.

\begin{lemma}[Multi-Packet Sufficient Statistic]\label{4lem_ss_mp}	
	Consider the observations $\bW_{1}$ and $\bW_{2}$ defined in \er{Ws_mp} and known training sequences in \er{Xs_def}. Then 
	\beq\label{Ys_MIMO_mp}
	\bY_{1} \ =\ \left(\frac{2N}{PT}\right) \bW_{1}\left(\bI_L\otimes\bX_1^H\right)\ , ~~\bY_2 \ = \ \left(\frac{2N}{PT}\right)\bW_{2}\left(\bI_L\otimes\bX_2^H\right)\ , 
	\eeq
	are sufficient for estimating unknown matrices $\bF$ and $\bSig_\bh$.  Moreover, $\bY_{1} - \bH$ and $\bY_{2} -\bF\bH$ are independent random matrices with i.i.d. $\mccn(0,\sigma^2)$ entries, where $\sigma^2\define 2N\sigma_n^2/PT$. 
	$\hfill\diamond$
\end{lemma}
\begin{proof}
	From the definition of $\bW_{1}$ in \er{Ws_mp} along with  $\bX_1$ in \er{Xs_def}, we can readily express $\bY_{1}$ in \er{Ys_MIMO_mp} as
	\beq
	\bY_1 \ = \ \bH + \left(\frac{2N}{PT}\right)\bN_1\left(\bI_L\otimes\bX_1^H\right) \ .\nn
	\eeq To show the entries of the last matrix are i.i.d., we vectorize it,
	\beq
	\left(\frac{2N}{PT}\right) \vc \left[\bN_1\cdot\left(\bI_L\otimes\bX_1^H\right)\right] \ = \  \left(\frac{2N}{PT}\right) \left[\left(\bI_L\otimes\bX_1^*\right)\otimes \bI_M\right]\vc \bN_1 \in\mbbC^{MNL} \ ,
	\eeq
	which is zero-mean and has covariance matrix  
	\beq
	\left(\frac{2N}{PT}\right)^2 \left(\bI_L\otimes\bX_1^*\bX_1^T\otimes \bI_M\right) \sigma_n^2\bI_{MNL} \ =\ \sigma^2\bI_{MNL} \ .\nn
	\eeq  
	Note  Kronecker product is associative and 
	$\vc \left(\bA\bB\bC\right) = \left(\bC^T\otimes\bA\right)\vc \bB$ is used \cite{brew}. This shows that $\bY_1 - \bH$ is a random matrices with i.i.d. $\mccn(0,\sigma^2)$ entries. Similarly, we can show that 
	\beq
	\bY_2 -\bF\bH \ = \ \left(\frac{2N}{PT}\right)\bN_2\left(\bI_L\otimes\bX_2^H\right) \ ,\nn
	\eeq 
	which is also a random matrices with i.i.d. $\mccn(0,\sigma^2)$ entries. The independence between these two matrices follow from that noises are independent over time and across packets \er{4observations_mp}. 
	
	From the Neyman-Fisher theorem \cite[pg. 117]{kay}, to prove sufficiency of \er{Ys_MIMO_mp} it suffices to show that $p\left(\bW_1,\bW_2;\bF,\bSig_\bh\right)$ factors into a product $g\left(\bY_1,\bY_2, \bF,\bSig_\bh\right) f\left(\bW_1,\bW_2\right)$, where $f$ does not depend on $\bY_1,\bY_2, \bF,\bSig_\bh$ and $g$ does not depend on $\bW_1,\bW_2$. 
	We prove this using the conditional pdf
	\beq
	p\left(\bW_1,\bW_2;\bF,\bSig_\bh\right) \ = \ E_\bH\left[p\left(\bW_1,\bW_2|\bH;\bF,\bSig_\bh\right)\right]\ ,
	\eeq
	where the expectation $E_\bH[\cdot]$ is with respect to $\bH$ as defined in \eqref{eq_Hmp_def}. Since $\bW_1$ and $\bW_2$ are conditionally independent given $\bH$, we have
	\bea\label{4lem_ss_mp_pdf}
	&&(\pi\sigma_n^2)^{MLT} p\left(\bW_1,\bW_2;\bF,\bSig_\bh\right) \nn\\ 
	&=& E_\bH\left[\exp\left(-\frac{1}{\sigma_n^2}\left\lVert\bW_1-\bH\left(\bI_L\otimes\bX_1\right)\right\rVert^2-\frac{1}{\sigma_n^2}\left\lVert\bW_2-\bF\bH\left(\bI_L\otimes\bX_2\right)\right\rVert^2\right)\right]\nn\\
%	&=&E_\bH\left[\exp\left(\frac{\Trace\left[\bW_1^H\bH\left(\bI_L\otimes\bX_1\right)\right]}{\sigma_n^2}+\frac{\Trace\left[\left(\bI_L\otimes\bX_1^H\right)\bH^H\bW_1\right]}{\sigma_n^2}\right. \right.\nn\\
%	&&\left. \left. -\frac{\Trace\left[\left(\bI_L\otimes\bX_1^H\right)\bH^H\bH\left(\bI_L\otimes\bX_1\right)\right]}{\sigma_n^2}+ \frac{\Trace\left[\bW_2^H\bF\bH\left(\bI_L\otimes\bX_2\right)\right]}{\sigma_n^2}\right. \right.\nn\\
%	&&\left. \left.+ \frac{\Trace\left[\left(\bI_L\otimes\bX_2^H\right)\bH^H\bF^H\bW_2\right]}{\sigma_n^2} -\frac{\Trace\left[\left(\bI_L\otimes\bX_2^H\right)\bH^H\bF^H\bF\bH\left(\bI_L\otimes\bX_2\right)\right]}{\sigma_n^2}\right)\right]\nn\\
%	&&\exp\left(-\frac{1}{\sigma_n^2}\left\lVert\bW_1\right\rVert^2-\frac{1}{\sigma_n^2}\left\lVert\bW_2\right\rVert^2\right) \nn\\
	&=& E_\bH\left[\exp\left(\frac{2\Real\Trace\left[\bH^H\bW_1\left(\bI_L\otimes\bX_1^H\right)\right]}{\sigma_n^2}-\frac{\Trace\left[\bH^H\bH\left(\bI_L\otimes\bX_1\bX_1^H\right)\right]}{\sigma_n^2}\right. \right.\nn\\
	&& \left. \left. + \frac{2\Real\Trace\left[\bH^H\bF^H\bW_2\left(\bI_L\otimes\bX_2^H\right)\right]}{\sigma_n^2} -\frac{\Trace\left[\bH^H\bF^H\bF\bH\left(\bI_L\otimes\bX_2\bX_2^H\right)\right]}{\sigma_n^2}\right)\right]\nn\\
	&&\exp\left(-\frac{1}{\sigma_n^2}\left\lVert\bW_1\right\rVert^2-\frac{1}{\sigma_n^2}\left\lVert\bW_2\right\rVert^2\right) \nn\\
	&=&E_\bH\left[\exp\left(\frac{2\Real\Trace\left[\bH^H\bY_1+\bH^H\bF^H\bY_2\right]}{\sigma^2}-\frac{\Trace\left[\bH^H\bH+\bH^H\bF^H\bF\bH\right]}{\sigma^2}\right) \right]\nn\\
	&&\exp\left(-\frac{1}{\sigma_n^2}\left\lVert\bW_1\right\rVert^2-\frac{1}{\sigma_n^2}\left\lVert\bW_2\right\rVert^2\right)\ ,
	\eea
	where $\lVert\bA \rVert^2=\Trace[\bA^H\bA]$ denotes the Frobenius norm. Also, the third equality follows from the identities $2\Real\Trace[\bA] = \Trace[\bA] + \Trace[\bA^H]$ and $\Trace[\bA\bB] = \Trace[\bB\bA]$, and the fourth equality follows from \er{Xs_def} and the definition of $\sigma^2$. 
	
	In the final expression of \er{4lem_ss_mp_pdf}, we denote the first factor by $(\pi\sigma_n^2)^{MLT}g\left(\bY_1,\bY_2, \bF,\bSig_\bh\right)$ and the second by $ f\left(\bW_1,\bW_2\right)$. Note through the expectation over $\bH$,  $g$ only depends on 
	$\bY_1,\bY_2, \bF, \bC_\bh$ (which is assumed known), and $\bSig_\bh$ but not on $\bW_1,\bW_2$. And $f$ only depends on $\bW_1,\bW_2$, but not $\bY_1,\bY_2, \bF,\bSig_\bh$. Thus, the Neyman-Fisher theorem applies\cite[pg. 117]{kay}, and this completes the proof. 
\end{proof}

Note $\bY_1$ and $\bY_2$ in \er{Ys_MIMO_mp} are a sufficient statistic regardless what the correlation matrix $\bC_\bh$ is. But $\bC_\bh$ will play a role in the PDF after the expectation over $\bH$. 
As in the last section, our ultimate goal is to find the maximum-likelihood (ML) estimators for $\btt$, 
\beq
\btt \ \define \vc \begin{bmatrix}
	\bF & \bSig_\bh
\end{bmatrix} \ ,
\eeq
where $\bF$ is defined in \er{F_MIMO} and $\bSig_\bh$ in \er{Sig_h_MIMO}. 
Using the multi-packet sufficient statistics in \er{Ys_MIMO_mp}, the multi-packet ML estimators for $\btt$ shall satisfy the following optimal criteria, 
\beq\label{4mle_criteria_mp}
\hat{\btt}_{ML} \ \define \ \arg\max_{\btt} p\left(\bY_1,\bY_2;\btt\right) \ .
\eeq
However, as we learned from its prequel \cite{wu_PCA}, these ML estimators
are unlikely in closed-form in general. Thus, we defer discussion on the ML estimators but first seek another set of estimators, i.e., the \textit{method of moments} estimators \cite[Ch. 9]{kay}. 

\begin{lemma}[Method of Moments Estimators]
	Let $\bY_1$ and $\bY_2$ be the sufficient statistics in \er{Ys_MIMO_mp}. Suppose $\bF$ and $\bSig_\bh$ are unknown. Consider the sample covariance matrix,
	\beq\label{S_mp_MIMO}
	\bS_{mp} \ \define \ \frac{1}{NL} \begin{bmatrix}
		\bY_1\bY_1^H &  \bY_1\bY_2^H\\
		\bY_2\bY_1^H & \bY_2\bY_2^H
	\end{bmatrix} \in\mbbC^{2M\times 2M}\ . 
	\eeq
	The eigen-decomposition of $\bS$ can be written as
	\beq
	\bS_{mp} \bU_\bs \ = \ \bU_\bs \diag(\mu_1,\dots,\mu_{2M}) \ ,
	\eeq
	where $\diag(\cdot)$ denotes a square diagonal matrix with its input as diagonal entries, and the eigen-values $\mu_k\geq 0$ are in descending order. Define the unitary eigen-vector matrix $\bU_\bs$ as a 2 by 2 block matrix, i.e., 
	\beq
	\bU_\bs \ \define \ \begin{bmatrix}
		\bU_{\bs11} & \bU_{\bs12} \\
		\bU_{\bs21} & \bU_{\bs22}
	\end{bmatrix} \ , 
	\eeq
	where $\bU_{\bs ij}\in\mbbC^{M\times M}$ and $i,j=1,2$. 
	Then, the method of moments (MM) estimators of $\bF$ and $\bSig_\bh$ are, respectively, conditioned on $\bU_{\bs11}$ is non-singular, 
	\beq\label{MME_ch4}
	\hat{\bF}_{MM}  \ = \  \bU_{\bs21} \bU_{\bs11}^{-1} \ , ~~~~ \hat{\bSig_\bh}_{MM} \ = \  \bU_{\bs11}\left(\diag(\mu_1,\dots,\mu_M) - \sigma^2\bI_M\right)^+ \bU_{\bs11}^H \ ,
	\eeq
	where $\sigma^2\define 2N\sigma_n^2/PT$ and the function $(\cdot)^+$ is defined in Theorem \ref{MLE_ch4}. 
\end{lemma}
\begin{proof}
The multi-packet sufficient statistics can be collected as 
\beq\label{Vs_MIMO}
\bV_s \ \define \ \begin{bmatrix}
	\bY_{1}\\
	\bY_{2} 
\end{bmatrix} \ = \ \begin{bmatrix}
\bH\\
\bF\bH
\end{bmatrix} + \bN_s\in\mbbC^{2M\times NL} \ ,
\eeq
where the noise is i.i.d., that is $\vc \bN_s\sim\mccn\left(\bzro_{2MNL}, \sigma^2\bI_{2MNL}\right)$, as proven in Lemma \ref{4lem_ss_mp}. 	
It is straightforward to show that
\beq
E\left[\bS_{mp}\right] \ = \ \frac{1}{NL}E\left[\bV_s\bV_s^H\right] \ = \ \begin{bmatrix}
	\bSig_\bh + \sigma^2\bI_M& \bSig_\bh \bF^H\\
	\bF\bSig_\bh & \bF\bSig_\bh\bF^H + \sigma^2\bI_M
\end{bmatrix}\ .
\eeq
Apparently,  the second moments of sufficient statistics $\bY_1$ and $\bY_2$ are functions of unknown parameters $\btt$, or $\bF$ and $\bSig_\bh$ \er{theta_MIMO}, 
\beq
E[\bS_{mp}] \ = \ \bT\left(\btt\right) \ = \ \begin{bmatrix}
	\bI_M\\
	\bF
\end{bmatrix} \bSig_\bh \begin{bmatrix}
\bI_M&\bF^H
\end{bmatrix} + \sigma^2\bI_{2M}\ , 
\eeq
where $\bT:\mbbC^{2M^2}\rightarrow\mbbC^{2M\times 2M}$ denotes a mapping. 
Then, from basic principles of MM estimation \cite[Sec.~9.4]{kay}, we find $\hat{\btt}_{MM}$ by the inverse of aforementioned mapping, 
\beq
\hat{\btt}_{MM} \ = \ \bT^{-1}\left(\bS_{mp}\right) \ .
\eeq
The formula of MM estimators in \er{MME_ch4} follows directly from the proof of Theorem \ref{MLE_ch4}. This completes the proof. 
\end{proof}

The MM estimators are provably consistent, easy to determine, and does not require knowing $\bC_\bh$. However, they are generally sub-optimal to the corresponding ML estimators \cite[Ch.~9]{kay}. Next we show a special case where the ML estimators and the MM estimators coincide, and then discuss how to find the ML estimators in general fading conditions if $\bC_\bh$ is known. 

\begin{corollary}[ML Estimators for Fast Fading]
	If the Rayleigh fading channel is temporally i.i.d., that is $\bC_\bh=\bI_L$ in \er{vec_Hmp},  then the multi-packet MM estimators given in \er{MME_ch4} for $\bF$ and $\bSig_\bh$ satisfy \er{4mle_criteria_mp} and hence are the maximum-likelihood estimators. 
\end{corollary}
\begin{proof}
	Conditioned on $\bC_\bh=\bI_L$,  the channel matrix $\bH$ in \er{vec_Hmp} would satisfy, 
	\beq
	\vc \bH \sim\mccn\left(\bzro_{MNL}, \bI_{NL} \otimes\bSig_\bh\right) \ .
	\eeq 
	Note the similarity between the distribution of this multi-packet $\bH$ and that of the single-packet channel matrix in \er{pdf_H_MIMO}. The proof follows Theorem \ref{MLE_ch4}, except the number of i.i.d. columns in $\bH$ is $NL$ rather $N$. 
\end{proof}

Next we discuss finding the ML estimator $\hat{\btt}_{ML}$ under general fading conditions. 
We assume the temporal correlation $\bC_\bh$ is known, and $L$ packets can be decorrelated by its eigen-vector matrix $\bQ$, i.e., 
\beq\label{Ch_eig_ch4}
\bQ^H\bC_\bh\bQ \ = \ \bD  \ \define \ \diag(d_1,\dots,d_L) \ . 
\eeq 
Note $\bC_\bh$ is normalized with 1's on its diagonal, i.e., $\Trace[\bC_\bh] = \Trace[\bD] =L$. 
Consider the decorrelated observation, i.e,.
\beq\label{V_mp}
\bV \ \define  \   \bV_s \left(\bQ^*\otimes \bI_N\right)  \ = \ \begin{bmatrix}
\bH_{d} \\
\bF\bH_{d}
\end{bmatrix} + \bN \in\mbbC^{2M\times NL}\ ,
\eeq
where $\vc \bN = [\left(\bQ^*\otimes \bI_N\right)^T\otimes\bI_{2M}]\vc\bN_s\sim\mccn\left(\bzro_{2MNL}, \sigma^2\bI_{2MNL}\right)$ is i.i.d. and,
\beq
\vc\bH_{d}\sim\mccn\left(\bzro_{MNL}, \bD \otimes\bI_N \otimes\bSig_\bh\right) .
\eeq
The understand this, for each $1\leq k\leq L$, we have $N$ i.i.d. complex Gaussian random vectors that follow $\mccn\left(\bzro_M, d_k\bSig_\bh\right)$, where $d_k$ are defined in \er{Ch_eig_ch4}.  

We define the log-likelihood function based on the pdf of $\bV\define [\bv_1, \dots, \bv_{NL}]$ is, 
\bea\label{pdf_mp_ch4}
\mathcal{L}(\btt)  & \define & \ln p\left(\btt;\bV\right)p(\bV; \btt) \nn \\
& = & C - N \sum_{k=1}^{L} \left(\ln \det \bSig_k + \sum_{i=1}^{N}\bv_{(k-1)N+i}^H\bSig_k^{-1}\bv_{(k-1)N+i}\right) \nn\\
& = & C - N \sum_{k=1}^{L} \left(\ln \det \bSig_k+\Trace[\bS_k\bSig_k^{-1}]\right)  \ ,
\eea
where $C$ is a constant independent of $\btt$ and we define for $1\leq k \leq L$, 
\beq
\bS_k \ \define \ \frac{1}{N}\sum_{i=1}^{N} \bv_{(k-1)N+i}\bv_{(k-1)N+i}^H \ , 
\eeq
and
\bea\label{CovYk}
\bSig_k  & \define & \begin{bmatrix}
	d_k\bSig_\bh + \sigma^2\bI_M& d_k\bSig_\bh \bF^H\\
	d_k\bF\bSig_\bh & d_k\bF\bSig_\bh\bF^H + \sigma^2\bI_M
\end{bmatrix} \nn\\
&=& d_k 
\begin{bmatrix}
  \bI_M\\
  \bF
\end{bmatrix} \bSig_\bh \begin{bmatrix}
  \bI_M&\bF^H
\end{bmatrix} + \sigma^2\bI_{2M} \ = \ \bB_1\bD_{1,k}\bB_1^H  + \bB_2\bD_2\bB_2^H \ , 
\eea
where $\bD_{1,k}\in\mbbC^{M\times M}$ and $\bB_i\in\mbbC^{2M\times M}$ (for $i=1,2$) are the block eigenvalues and orthonormal block eigenvectors of $\bSig$, 
\bea\label{eq_Dk}
\bB_1 &=&
\begin{bmatrix}
  \bI_M\\
  \bF
\end{bmatrix} \bA_1^{-\frac{1}{2}} \ , ~~~ \bB_2 \ = \ \begin{bmatrix}
  -\bF^H\\
  \bI_M
\end{bmatrix} \bA_2^{-\frac{1}{2}} \ ,  \nn\\
\bD_{1,k} &=& d_k \bA_1^{\frac{1}{2}}\bSig_\bh\bA_1^{\frac{1}{2}} + \sigma^2\bI_M \ , ~~~ \bD_2 \ = \ \sigma^2\bI_M \ , 
\eea
and $\bA_1$ and $\bA_2$ are positive-definite matrix defined in \er{As}. 
Since eigenvalues of $\bD_{1,k}$ are also eigenvalues of $\bSig$ \cite[Th. 1.1]{pere}, and observe that $\bD_2$ is already diagonal, we write down the (scalar) eigenvalue decomposition of $\bSig$,
\beq\label{Sig_eig_mp}
\bSig_k \ = \  \bU \begin{bmatrix}
  \bLam_{1,k} & \bzro_{M\times M}  \\
  \bzro_{M\times M} & \sigma^2\bI_M 
\end{bmatrix} \bU^H \ \ ,
\eeq
where we define 
\beq\label{eq_U_mp}
\bU \ \define \ \begin{bmatrix}
\bB_1 \bQ & \bB_2
\end{bmatrix} \ ,
\eeq
and we have 
\bea
\bD_{1,k}\bQ &=& \bQ\bLam_{1,k} \ , \nn\\
\bLam_{1,k}&=&\diag(\lambda_{1,k},\dots,\lambda_{M,k}) \ ,
\eea
 with $\bQ\in\mbbC^{M\times M}$ unitary. Note $\lambda_{j,k} \geq \sigma^2$ for all $1\leq j \leq M$ and $1\leq k\leq L$. In other words, the covariance matrix $\bSig_k$ has $2M$ real eigenvalues, where the largest $M$ eigenvalues are greater than or equal to $\sigma^2$, while the smallest $M$ eigenvalues are exactly  $\sigma^2$.
 
 Note that quality of virtual channels ($d_k$) after decorrelation is buried in $\lambda_{j,k}$.  Because of the freedom of $d_k$ as in definition of $\bD_{1,k}$ in \er{eq_Dk}, a closed-form expression for the ML estimator is mathematically intractable. Since seeking a fast algorithm that allows real-time impedance estimation is our goal, we consider numerical methods based on iterations (e.g., gradient descent or Newton's method) are out of the scope of this paper. Instead, we use the fundamental lower bound in CRB as a reference when we evaluate the ML or MM estimators in the numeral section.

%In order to find the ML estimators, we define the log-likelihood function as
%\bea
%\mathcal{L}(\btt) &\define& \ln p\left(\btt;\bV\right) \ = \ \ln \left[\prod_{k=1}^{L}\det(\pi \bSig_k)^{-N} \cdot \exp \left(-N\Trace[\bS_k\bSig_k^{-1}]\right) \right]\nn\\
%& = & C - N\sum_{k=1}^L\left( \ln \det \bSig_k   +\Trace[\bS_k\bSig_k^{-1}]\right) \ ,
%\eea
%where $C$ is a constant independent of $\btt$. A necessary condition for joint ML estimators is that the complex gradient  vanishes, i.e.,
%\beq
%\frac{\partial \mathcal{L}(\btt)}{\partial \btt^*} \ = \ \bzro \ .
%\eeq
%However, it is unclear if a closed-form solution to this condition exists. Thus, one could use numerical methods to find these  ML estimators. In particular, a variation of the
%complex Newton-Raphson\footnote{This is called the quasi-Newton method in a  tutorial on Wirtinger calculus (or $\mbbC\mbbR$ calculus) \cite{kreu}. } iteration can be used to find the true MLE\cite[eq. 11]{yan},
%\beq\label{newton_ch4}
%{\btt}_{p+1} \ = \ {\btt}_p - \mathcal{H}^{-1}(\btt_p)  \left.\frac{\partial \mathcal{L}(\btt)}{\partial \btt^*}\right|_{\btt=\btt_p} \ ,
%\eeq
%where the complex Hessian of the log-likelihood function is,
%\beq\label{Hessian_mp}
%\mathcal{H}(\btt) \ \define \ \frac{\partial^2 \mathcal{L}(\btt)}{\partial \btt^*\partial\btt^T}  \ . 
%\eeq

The multi-packet FIM follows directly from \eqref{FIM_MIMO}, 
\beq\label{FIMmp}
[\boldsymbol{\mathcal{I}}_{mp}(\btt)]_{ij} \ = \ N\cdot \sum_{k=1}^{L}\Trace\left[\bSig_k^{-1}\frac{\partial \bSig_k}{\partial \theta_i^*}\bSig_k^{-1}\frac{\partial \bSig_k}{\partial \theta_j}\right] \ , 
\eeq
where $\bSig_k$ is the $k$-th covariance matrix defined in \er{CovYk}. Similarly, the error covariance matrix $\bC_{\hat{\btt}}$ of any unbiased estimator $\hat{\btt}$ is lower bounded by the Cram\'er-Rao bound (CRB), which is  the inverse of $\boldsymbol{\mathcal{I}}_{mp}(\btt)$,
\beq\label{crb_mp_ch4}
\bC_{\hat{\btt}} \ \define \ E_{\bY_1,\bY_2;\btt}\left[\left(\hat{\btt}-\btt\right)\left(\hat{\btt}-\btt\right)^H\right] \ \geq \ \boldsymbol{\mathcal{I}}_{mp}(\btt)^{-1}\ , 
\eeq
where the expectation is with respect to pdf in \er{pdf_mp_ch4} and $\bA\geq \bB$ means $\bA-\bB$ is positive semi-definite.

%The multiple CRB is scaled by $1/L$, yet the multi-packet MCB becomes, conditioned on $NL> M$, 
%\beq\label{MCB_mp}
%\mathcal{M}_{mp}(\bF) \ = \ \frac{\sigma^2 }{NL-M}\Trace\left(\bSig_\bh^{-1}\right) \Trace \left(\bF\bF^H + \bI_M\right) \ . 
%\eeq

For any estimator of $\bF$, we find an estimator for antenna impedance via \er{F_MIMO}, i.e.,
\beq\label{zAhat}
\hat{\bZ}_\bA \ = \ \left(\bZ_1-\bZ_2\bR_2^{-1/2}\hat{\bF}_{MM}\bR_1^{1/2}\right)\left(\bR_2^{-1/2}\hat{\bF}_{MM}\bR_1^{1/2}-\bI_M\right)^{-1}\ . 
\eeq
However, due to the reciprocity theorem of electromagnetics\cite[pg. 144]{bala}, $\bZ_\bA$ is symmetric and so should any reasonable estimate of it. Here we replace $\hat{\bZ}_\bA$ by its nearest symmetric matrix, i.e.,  
\beq\label{zA_est_MIMO}
\tilde{\bZ}_\bA \ \define \ \frac{1}{2}\left(\hat{\bZ}_\bA + \hat{\bZ}_\bA^T \right) \ .
\eeq

Based on this new estimate, the receiver matches its load impedance for minimum noise-figure, which reduces to maximum power transfer under our noise model \cite[eq.~10]{deso}, $\hat{\bZ}_L =  \tilde{\bZ}_\bA^*$. 
Consequently, we calculate an excess (transmit) power needed for this matching compared to the truly optimal one, i.e., $\bZ_{L,opt} = \bZ_\bA^*$,
\beq\label{ex_pwr}
10\log_{10}\left(\Trace\left[\left(4\bR_\bA\right)^{-1}\bSig_\bg\right]/E\Trace\left[\left(\tilde{\bZ}_\bA^*+\bZ_\bA\right)^{-H}\tilde{\bR}_\bA\left(\tilde{\bZ}_\bA^*+\bZ_\bA\right)^{-1}\bSig_\bg\right] \right) \ ,
\eeq
where $\tilde{\bR}_\bA \define \Real\{\tilde{\bZ}_\bA\}$.

In the next section,  we compare the performance of  estimators derived in this paper against their corresponding lower bounds, and explore the potential benefits of these estimators on system-level metrics, such as channel capacity.

\section{Numerical Results}\label{4secV}
In this section, we explore the performance of  estimators in the previous section through numerical examples. Consider a narrow-band MIMO communications system with $N=4$ transmit antennas and $M=2$ receive antennas, whose carrier frequency is 2.1 GHz. This frequency is chosen based on the first E-UTRA down-link operating band in LTE specifications\cite{3gpp_TS36101}. The duration of each data packet equals to a subframe of LTE, i.e., $T_s=1$ ms. Block-fading channel is assumed, such that during one data packet, the channel information remains the same, but it generally varies from packet to packet \cite{bigu}. 

For each data packet, a training sequence precedes data sequence \cite[Fig. 1(a)]{liu}. 
We take the two partitions of the training sequence $\bX=[\bX_1, \bX_2]$ in \er{Xs_def} from a normalized discrete Fourier transform
(DFT) matrix of dimension $K=T/2=32$, e.g., \cite[eq. 10]{bigu}. In particular, the first part $\bX_1$ is chosen as the first $N$ rows, while $\bX_2$ the next $N$ rows, and $\bX_i\bX_i^H=K\bI_N$ for $i=1,2$.  The unknown antenna impedance is that of a uniform linear array (ULA) \cite{domi2}, i.e., 
\beq
\bZ_\bA \ = \  \begin{bmatrix}
	72.8521 + j1.6869 &-15.7457 -j27.8393\\
	-15.7457 -j27.8393 & 72.8521 + j1.6869
\end{bmatrix} \,\Omega \ .
\eeq The load impedance is $\bZ_1=50\bI_M \,\Omega$ for the first $K=32$ symbols of each training sequence, and $\bZ_2=(50+j20)\bI_M+10\times{\bf 1} \,\Omega$ for the remaining $T-K=32$ symbols, where ${\bf 1}$ is the $M$ by $M$ all one matrix. From \er{F_MIMO}, it follows that 
\beq
\bF \ = \ \begin{bmatrix}
	0.9804 - j0.1613&   0.0261 - j0.0334\\
	0.0261 - j0.0334&  0.9804 - j0.1613
\end{bmatrix} \ .
\eeq

In this section, we explore important properties of the estimators derived in previous sections. 
The average post-detection SNR of a received symbol is defined from \er{sgl_model_MIMO} as \cite[Sec. VIII]{bigu},
\beq\label{symbolSNR}
\gamma \ \define \ \frac{E\Trace\left[\bH\bx\bx^H\bH^H\right]}{E\Trace\left[\bn_L\bn_L^H\right]} \ = \ \frac{P\sigma_H^2}{\sigma_{n}^2}\ ,
\eeq
where $\sigma_{n}^2$ is the noise variance at each port of the $M$-port receiver and $\sigma_H^2$ is the mean of diagonal entries of $\bSig_\bh$ in \er{pdf_H_MIMO}, 
\beq\label{sgmH2_MIMO_ch5}
\sigma_H^2 \ \define \ \frac{1}{M}\Trace[\bSig_\bh] \ .
\eeq

\begin{figure}[t!]
	\begin{center}
		\includegraphics[width=.6\textwidth, keepaspectratio=true]{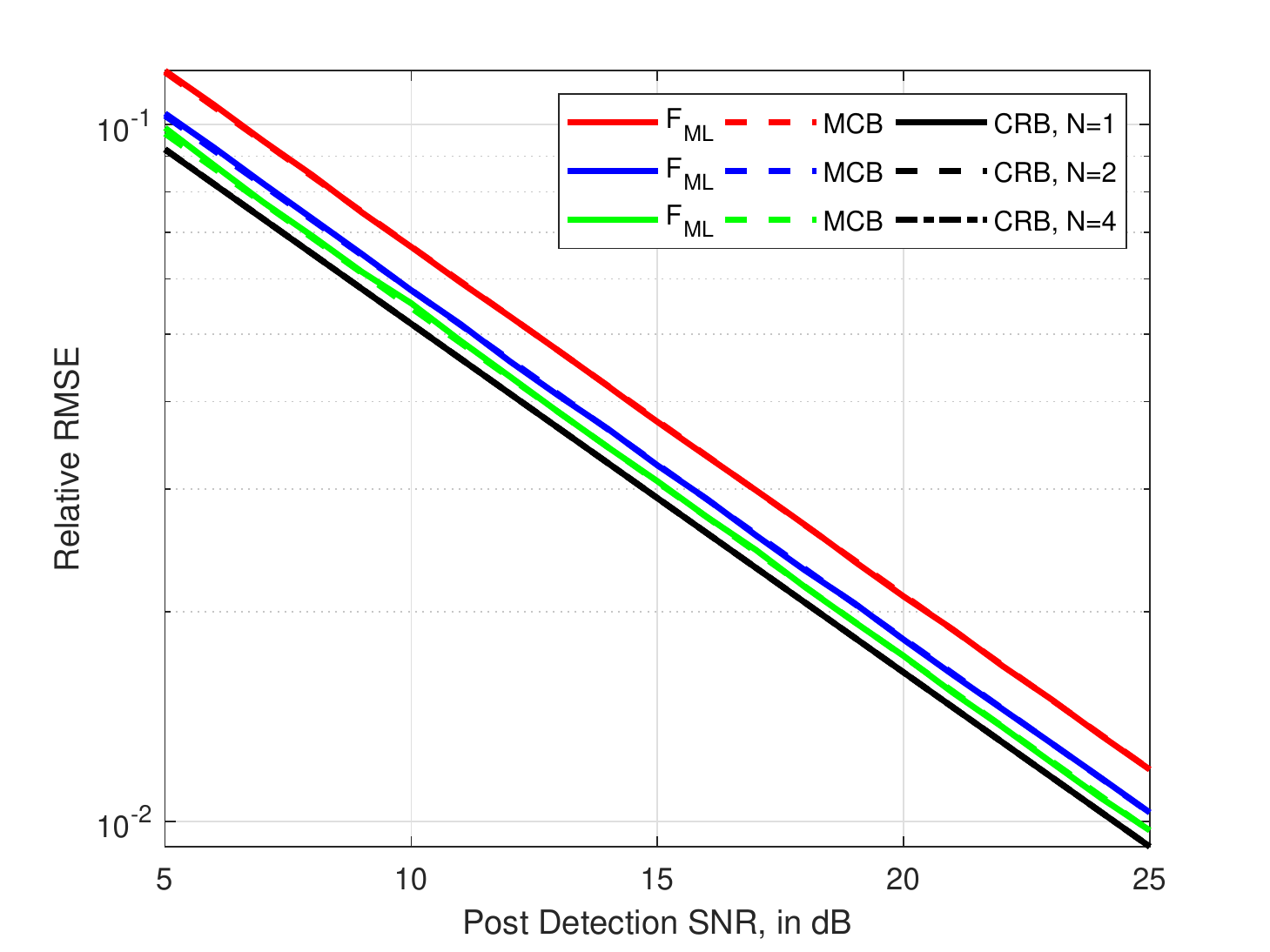}
	\end{center}
	\vspace{-12pt}
	\caption{Relative MSE of  $\hat{\bF}_{ML}$ versus SNR in i.i.d. Fading, $L=5$. }
	\vspace{-18pt}
	\label{fig_iid_F_MIMO}
\end{figure}

As shown is Fig. \ref{fig_iid_F_MIMO}, the relative root mean-square error (RMSE) is plotted against SNR \er{symbolSNR}.  The ML estimator in $\hat{\bF}_{ML}$, for a given $L=5$ packets,  becomes efficient as the number of transmit antenna increases, i.e., more spatial diversity. We also observe that the Miller-Chang bound (MCB) is tighter than the CRB and touches the RMSE for all values of $L$ and SNR plotted in Fig.~\ref{fig_iid_F_MIMO}. Although the ML estimators are asymptotically unbiased and efficient, i.e., it achieves its corresponding CRB, the MCB (if exists) better predicts the RMSE of $\hat{\bF}_{ML}$ for finite sample size in $L$.  For different $N$ the CRB generally have different values as indicated by the formulas of Fisher information matrix in \er{FIM_MIMO}, but their numerical evaluations
seem indistinguishable in Fig.~\ref{fig_iid_F_MIMO}.

Next, we investigate the performance of estimators derived previously under different Rayleigh fading conditions, i.e., fast, medium, and slow fading \cite{wu3}. In particular, Clarke's model is assumed \cite{badd,zhen} and the normalized channel correlation matrix is,
\beq
\bC_\bh \ = \ \begin{bmatrix}
	R[0] & R[-1] & \cdots & R[-L+1]\\
	R[1] & R[0] & \cdots& R[-L+2] \\
	\vdots & \ddots & \ddots & \vdots \\
	R[L-1] & R[L-2]& \cdots  & R[0]
\end{bmatrix} \ ,
\eeq
where   $R[l] \ = \ J_0(2\pi f_dT_s |l|)$, 
$J_0(\cdot)$ is the zeroth-order Bessel function of the first kind, $T_s=1~ms$  is the sampling interval, and $l$ is the sample difference. The fading frequency (maximum Doppler frequency) is 
$f_d\define v/\lambda$, 
where $v$ is the velocity  of the fasting moving scatterer and $\lambda$ the wave-length of the carrier frequency. 

%%%%%%%%%%%%%%%%%%%%%%%%%%%
%
%		simulations of RMSE and bias of 4 by 2 MIMO, L=10, 		  for all fadings. 
%
%%%%%%%%%%%%%%%%%%%%%%%%%%%
%\begin{figure}
%	%	\vspace{-16pt}
%	\centering
%	\begin{subfigure}{0.5\textwidth}
%		\centering
%		\includegraphics[width=1.05\textwidth, keepaspectratio=true]{./figs/MIMO_Fmm_MSE_L10_4by2_12_17_18}
%		%		\vspace{-24pt}
%		\caption{Relative MSE of  $\hat{\bF}_{MM}$.}
%		%		\vspace{-12pt}
%		\label{fig_Fmm_fd_4by2}
%	\end{subfigure}\hfill
%	\begin{subfigure}{0.5\textwidth}
%		\centering
%		\includegraphics[width=1.05\textwidth, keepaspectratio=true]{./figs/MIMO_Fmm_bias_L10_4by2_12_17_18}
%		%		\vspace{-24pt}
%		\caption{Relative Absolute Bias of  $\hat{\bF}_{MM}$.}
%		%		\vspace{-12pt}
%		\label{fig_bias_4by2}
%	\end{subfigure}
%	\caption{Properties of $\hat{\bF}_{MM}$ for a 4 by 2 MIMO, $L=10$}
%	\vspace{-18pt}
%\end{figure}

\begin{figure}[t!]
	\begin{center}
		\includegraphics[width=.6\textwidth, keepaspectratio=true]{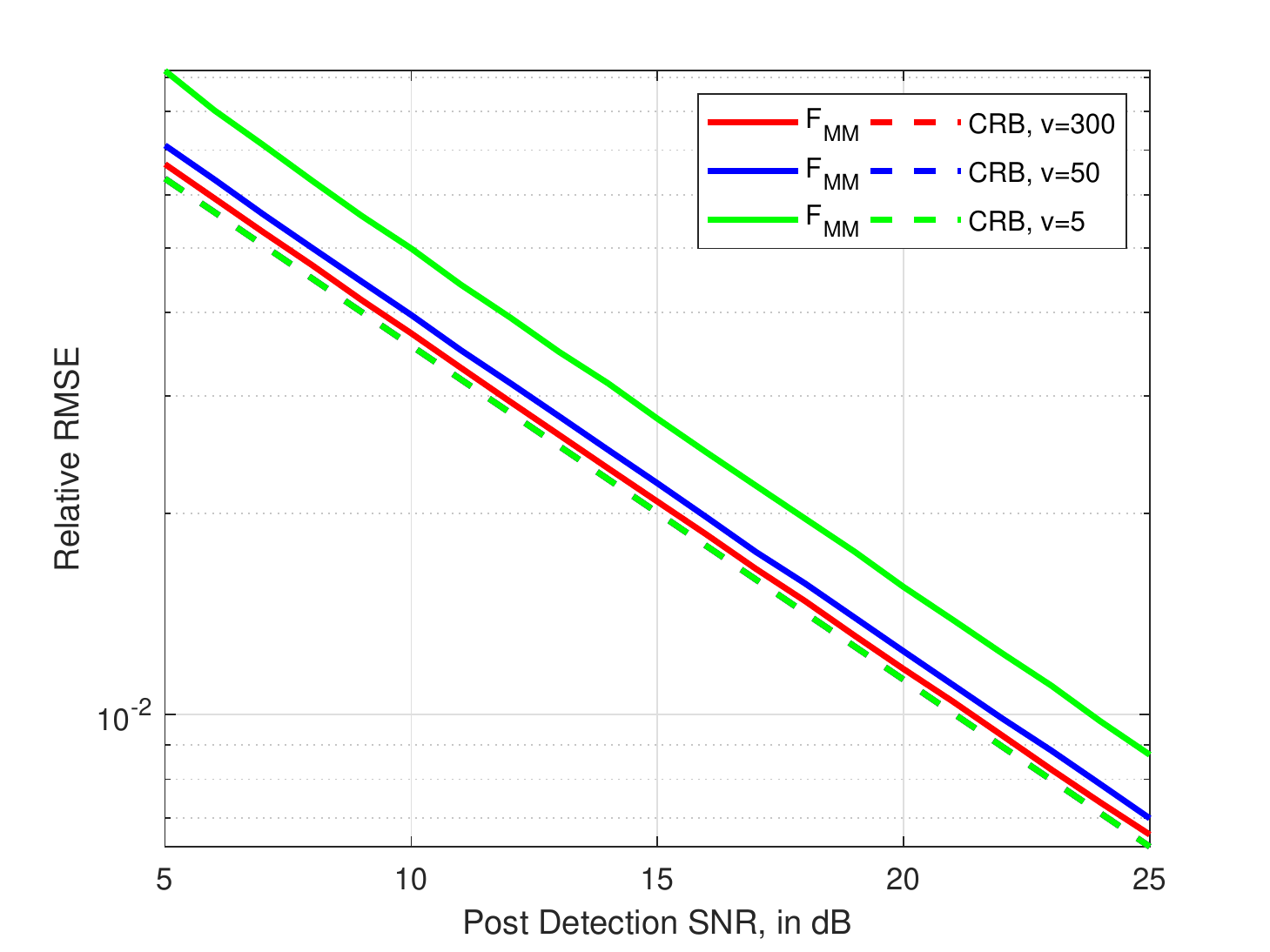}
	\end{center}
	\vspace{-12pt}
	\caption{Relative MSE of  $\hat{\bF}_{MM}$ for a 4 by 2 MIMO, $L=10$}
	\vspace{-18pt}
	\label{fig_Fmm_fd_4by2}
\end{figure}

In Fig.~\ref{fig_Fmm_fd_4by2}, we plot the relative RMSE of the method of moments (MM) estimators $\hat{\bF}_{MM}$ \er{MME_ch4}, for  a MIMO with $N=4$ transmit  and $M=2$  receiver antennas. The velocity of the fastest moving scatterer is $v=300, 50$ and $5$ km/h, which represents a fast, medium, and slow fading scenario, respectively. The MM estimator  $\hat{\bF}_{MM}$ is about 3 dB aways from its CRB under slow fading, and this gaps narrows to less 1 dB for medium and fast fading. Thus, a faster fading results in improved impedance estimation accuracy. This is reasonable as fast fading means more temporal diversity and less correlation between observations. Similar to Fig.~\ref{fig_iid_F_MIMO}, the CRB depends very little on fading conditions;  the CRB's of three cases considered are indistinguishable.

Next we evaluate the excess power defined in \er{ex_pwr}. A faster fading channel results in a smaller excess power. This means the transmitter may save power for an intended receive SNR, due to an improved match between antenna and load after impedance estimation and mismatch compensation. For example, the gain between the fast and slow fading cases is about 3 dB at low SNR. If a 0.5 dB excess power or less is considered a good match in practice, then it is achieved at relatively low SNR for all fading conditions. Further, the excess power vanishes  at high SNR. Next we give two examples which demonstrates the benefits of this impedance estimation algorithm in terms of  ergodic capacity.  

\begin{figure}[t!]
	\begin{center}
		\includegraphics[width=.6\textwidth, keepaspectratio=true]{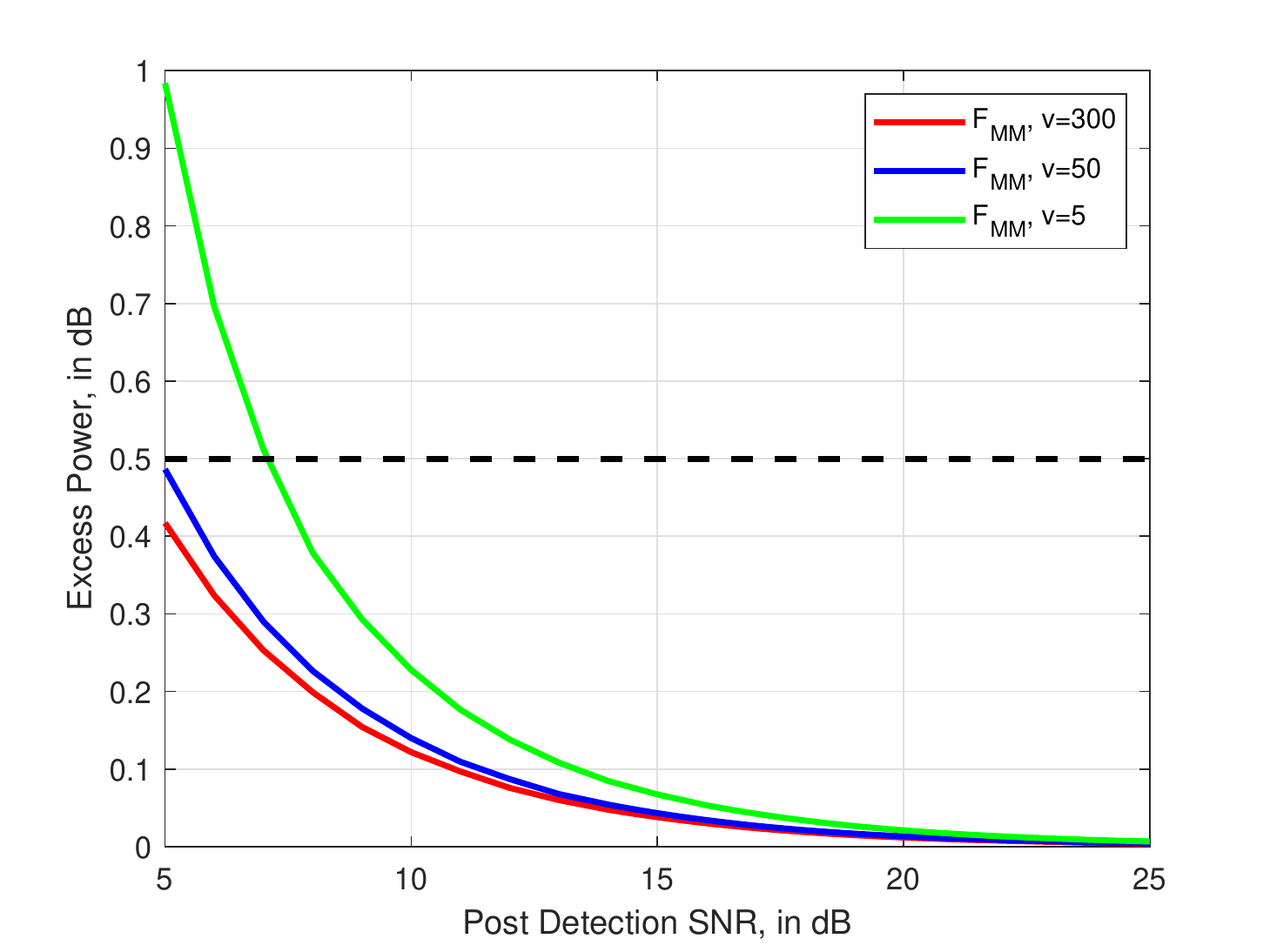}
	\end{center}
	\vspace{-12pt}
	\caption{Excess Power of  $\hat{\bF}_{MM}$ for 4 by 2 MIMO.}
	\vspace{-18pt}
	\label{fig_pwr_4by2}
\end{figure}

As derived by Hassibi and Hochwald, a lower bound on (ergodic) capacity exists, which incorporates the MMSE channel estimation error\cite[eq.~21]{Hassibi}, i.e.,
\beq
\ C_l \ = \ E \left[\log_2\det\left(\bI_M +\gamma_\text{eff}\cdot \frac{1}{N} \bSig_{\tilde{\bh}} \bH_w\bH_w^H\right)\right]  \ ,
\eeq
where  $\vc \bH_w\sim\mccn(\bzro, \bI_{MN})$, $\bSig_{\tilde{\bh}}$ is the normalized version of $\bSig_{\bh}$ such that $\Trace[\bSig_{\tilde{\bh}}]/M=1$, and with \er{symbolSNR} the effective SNR is defined as,
\beq
\gamma_\text{eff}  \ = \ \frac{P\sigma_{H}^2}{\sigma_{n}^2} \frac{PT\sigma_{H}^2}{PT\sigma_{H}^2+N(P\sigma_{H}^2+\sigma_{n}^2)} \ = \ \gamma \,\frac{1}{1+(1+1/\gamma)N/T}\ .
\eeq

%%%%%%%%%%%%%%%%%%%%%%%%%
%	Simulations for ergodic capacity vs SNR, one packet
%%%%%%%%%%%%%%%%%%%%%%%%%
\begin{figure}[t!]
    \begin{center}
        \includegraphics[width=.6\textwidth, keepaspectratio=true]{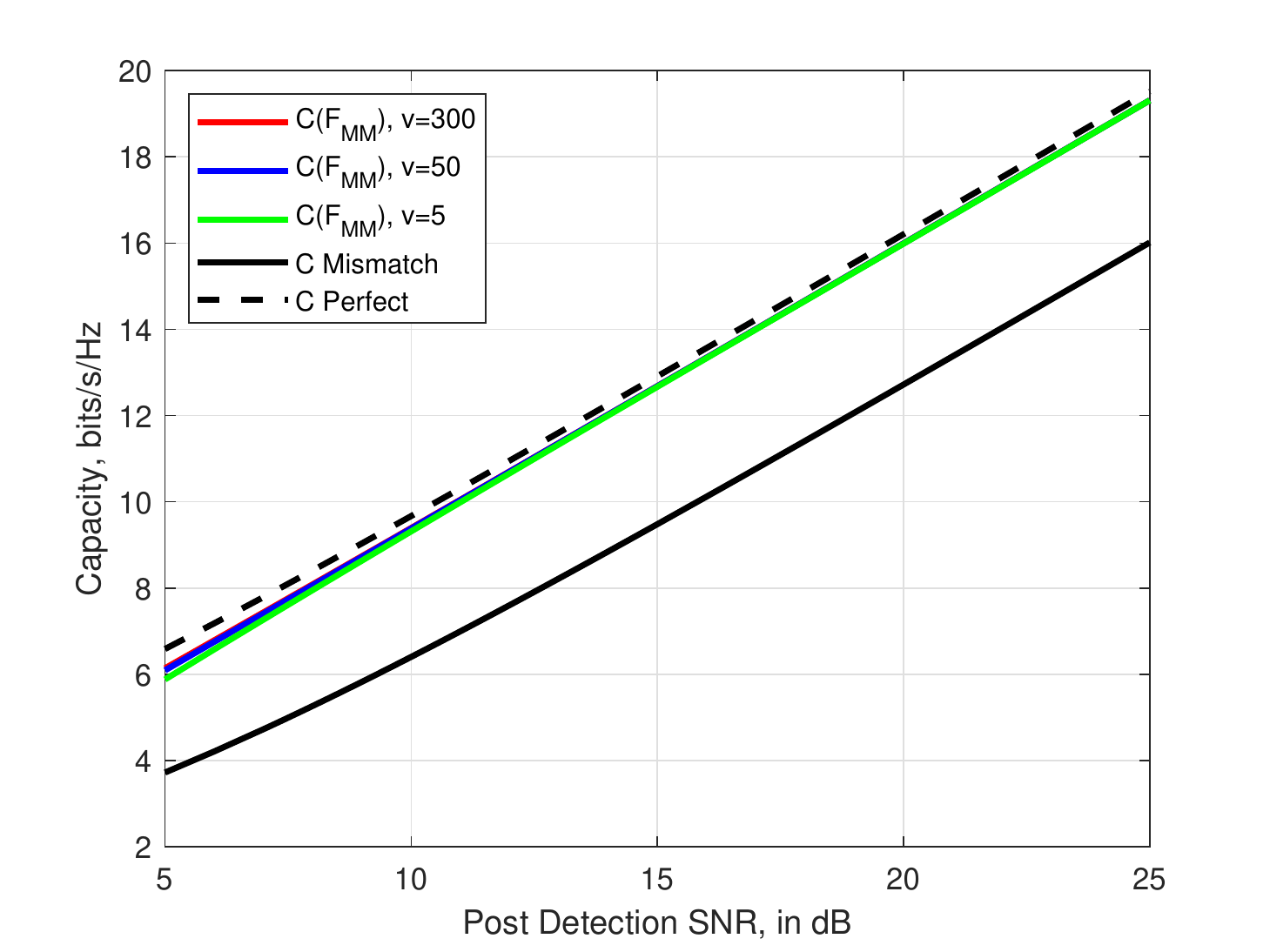}
    \end{center}
    \vspace{-12pt}
    \caption{Ergodic Capacity over SNR for 4 by 2 MIMO.}
    \vspace{-18pt}
    \label{fig_capacity_gain}
\end{figure}

Shin and Lee derived an upper bound for this ergodic capacity in closed-form, putting the expectation between $\log_2(\cdot)$ and $\det(\cdot)$ by Jensen's inequality \cite[Th.~III.2]{shin}, i.e.,
\beq
C_l \ \leq \ \log_2\left(\sum_{k=0}^{M} \left[\left(\frac{\gamma_\text{eff}}{N}\right)^kk! \,\sigma_k\left(\bSig_{\tilde{\bh}}\right)\cdot \sigma_k\left(\bI_N\right)\right]\right) \ ,
\eeq
where $M\leq N$ is assumed  and $\sigma_k(\bA)$ denotes the sum of all the $k$-rowed principal minor determinants of a square matrix $\bA$ \cite[pg. 17]{horn}. In particular, we have \cite[Th.~II.3]{shin},
\beq
\sigma_k\left(\bI_N\right) \ = \ {N \choose k} \ = \ \frac{N!}{k! (N-k)!} \ .
\eeq

Consider a 4 by 2 MIMO system again, i.e., $N=4$ and $M=2$. This ergodic capacity  upper bound boils down to
\beq\label{capacity_MIMO}
C_l \ \leq \ \log_2\left[1 + 2\gamma_\text{eff} + \frac{3}{4} \gamma_\text{eff}^2\cdot \det(\bSig_{\tilde{\bh}})\right] \ .
\eeq
Although calculating an upper bound instead of the exact ergodic capacity is less than ideal, it should qualitatively demonstrate the capacity boost using our proposed antenna impedance estimation algorithm.

In Fig.~\ref{fig_capacity_gain}, ergodic capacity upper bound \er{capacity_MIMO} are plotted against SNR of an originally mismatched receiver. The power loss due to mismatch is chosen as $5$dB. After applying our algorithm and matching to the estimate of $\bZ_\bA$ \er{zA_est_MIMO}, a significant gain on ergodic capacity $C(\hat{\bF}_{MM})$ is observed, compared to the mismatched receiver (the black solid line). This gain ranges from about 50\% at low SNR to 20\% at high SNR. The black dash line represents an upper bound on ergodic capacity, where the receiver is always optimally matched and observes the channel without errors. This upper bound, although unachievable by any practical system, is closed in by  $C(\hat{\bF}_{MM})$ to around 1 dB or less for all SNR and fading conditions considered. Also note at low SNR, faster fading leads to a marginally capacity boost, which vanishes  as SNR increases. 

%%%%%%%%%%%%%%%%%%%%%%%%%
%	Simulations for ergodic capacity vs d/lambda, one packet
%%%%%%%%%%%%%%%%%%%%%%%%%
\begin{figure}[t!]
	\begin{center}
		\includegraphics[width=.6\textwidth, keepaspectratio=true]{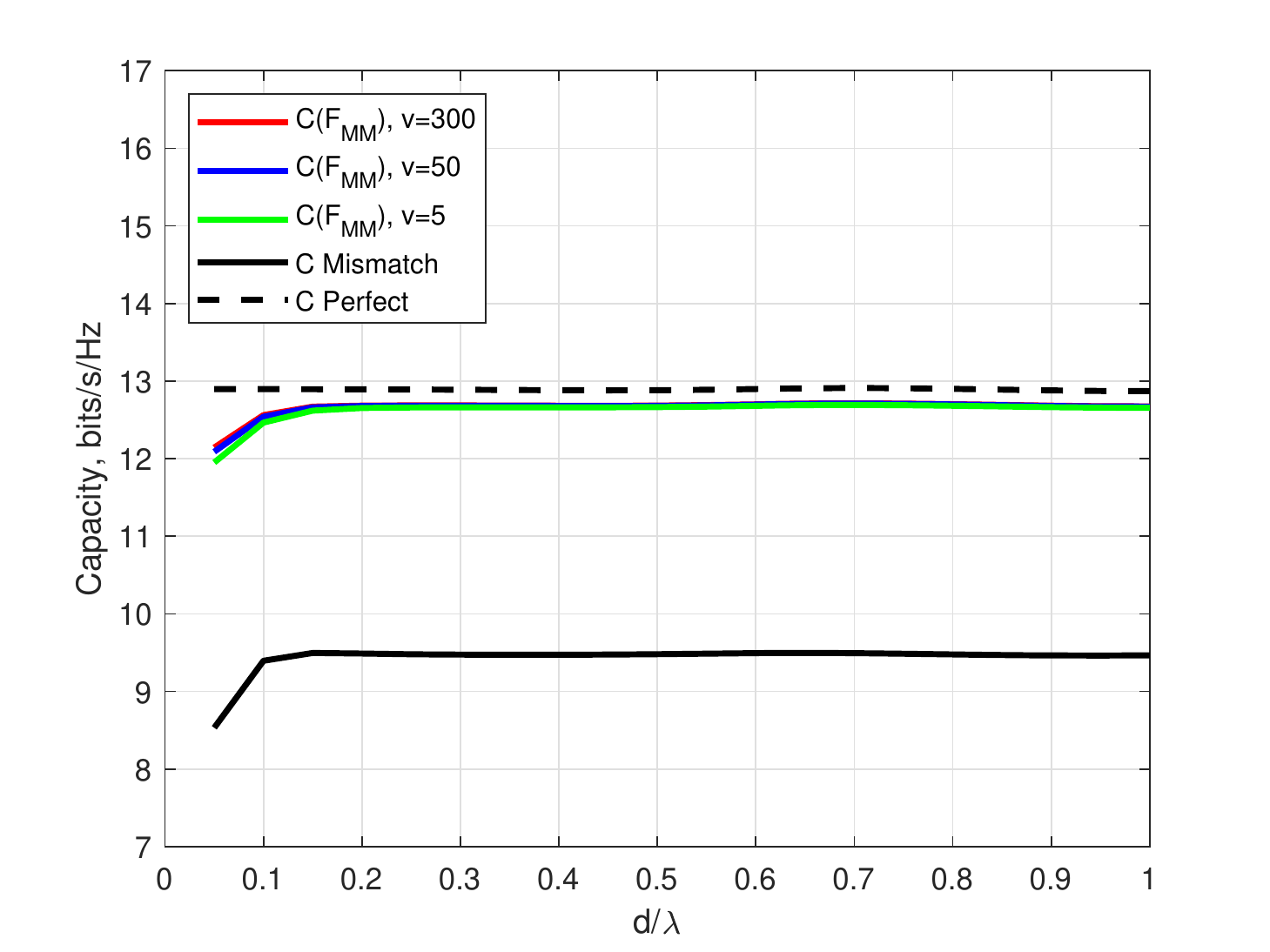}
	\end{center}
	\vspace{-12pt}
	\caption{Ergodic Capacity over $d/\lambda $ for 4 by 2 MIMO, SNR = 10dB.}
	\vspace{-18pt}
	\label{fig_capacity_gain_over_d}
\end{figure}

Plotted in Fig.~\ref{fig_capacity_gain_over_d} is  the ergodic capacity upper bound in \er{capacity_MIMO} versus antenna element-separation $d/\lambda$. The SNR for the originally mismatched receiver is fixed at 10 dB, while other settings remain identical as in Fig.~\ref{fig_capacity_gain}. Similar observations are also made here, as the $C(\hat{\bF}_{MM})$'s hone in the practically unachievable upper bound (the black dash line) within a fraction of 1 bit/s/Hz. This upper bound seems to depend very little on antenna spacing, yet the other capacity curves tend to drop for closely spaced arrays. Compared to the mismatched receiver, our algorithm improves capacity by over 30\% for all data points.

\section{Conclusion}\label{4secVI}
In this paper, we derived  the maximum-likelihood (ML) and method of moments (MM) estimators for MIMO  antenna impedance  using training sequences in various fading conditions. In particular, under i.i.d. fading, the ML estimator was derived as the ratio out of the top block eigen-vector of the sample covariance matrix. This ML estimator was shown to be a MM estimator under temporally correlated Rayleigh fading.  We also derived two fundamental lower bounds on these estimators, and explored the performance of these estimators through numerical examples. The ML and MM estimators become efficient (to CRB) when sufficient spatial and/or temporal diversity exists. A typical rule of thumb is the number of diversity is four times the number of receive antennas. Additionally,  trade-off between channel correlation and impedance estimation accuracy was investigated. Our numerical results indicate that the MIMO antenna impedance can be accurately estimated in a matter of milliseconds. This estimate is able to compensate power losses due to impedance mismatch partially at low SNR and almost all at high SNR. In the example of ergodic capacity, if the original mismatch at the receiver is significant, large capacity boost can be observed in general.

\bibliographystyle{unsrt}

\end{document}